\documentclass[pre,twocolumn,twoside,byrevtex,superscriptaddress,floatfix]{revtex4-1}

\usepackage{currfile}
\usepackage{graphicx,epsfig,verbatim,enumerate}
\usepackage{amssymb,amsmath}
\usepackage{ifthen}
\usepackage{subcaption}
\usepackage{subcaption}
 \usepackage{booktabs}
\usepackage{mathtools}
\usepackage{url}
\usepackage{hyperref}
\newboolean{twocolswitch}

\newcommand{\sindex}[1]{}
\newcommand{\nindex}[1]{}

\newcommand{\www}[1]{\url{#1}}

\usepackage{lettrine}

\usepackage{color}

\setboolean{twocolswitch}{true}

\begin{document}

\title{\protect
Global brain drain and gain in high-potential student mobility
}

\author{
  \firstname{Tabia Tanzin}
  \surname{Prama}
}

\email{tabia.prama@uvm.edu}

\affiliation{
  Computational Story Lab,
  Vermont Complex Systems Institute,
  Vermont Advanced Computing Center,
  University of Vermont,
  Burlington, VT 05405, US.
}

\author{
  \firstname{Christopher M.}
  \surname{Danforth}
}

\email{chris.danforth@uvm.edu}

\affiliation{
  Computational Story Lab,
  Vermont Complex Systems Institute,
  Vermont Advanced Computing Center,
  University of Vermont,
  Burlington, VT 05405, US.
}

\affiliation{
  Department of Mathematics and Statistics,
  Burlington, VT 05405, US.
}

\author{
  \firstname{Peter Sheridan}
  \surname{Dodds}
}

\email{peter.dodds@uvm.edu}

\affiliation{
  Computational Story Lab,
  Vermont Complex Systems Institute,
  Vermont Advanced Computing Center,
  University of Vermont,
  Burlington, VT 05405, US.
}

\affiliation{
  Department of Computer Science,
  Burlington, VT 05405, US.
}

\affiliation{
  Santa Fe Institute,
  1399 Hyde Park Rd,
  Santa Fe,
  NM 87501,
  US
}
\date{\today}

\begin{abstract}

The mobility of high-potential individuals, particularly graduates from elite academic institutions, serves as a critical driver of global innovation and economic development. Despite its importance, granular data on the specific trajectories and demographic drivers of these flows remain scarce in traditional administrative sources. In this study, we leverage anonymized, aggregate-level digital trace data from the LinkedIn Advertising platform to map the international mobility of graduates from 1,504 QS-ranked universities across 102 countries. We find that global talent flows are highly concentrated, with the United States capturing 38.4\% of the mobile elite, followed by the United Kingdom (7.9\%) and Canada (6.8\%), while regional hubs like the United Arab Emirates (5.2\%) have emerged as significant talent magnets. Our analysis reveals a global Relative Gender Gap (RGG) of +3.16\%, indicating a modest male overrepresentation that varies sharply by destination, from extreme male skews in Ethiopia (+60.34\%) to female overrepresentation in Armenia ($-$30.77\%). Professional integration is highly structured; while Business Development and Operations are universal entry channels, technical specialization in Engineering and IT is concentrated in specific innovation hubs. Destination ``pull'' is primarily driven by economic capacity, institutional stability, and educational infrastructure, though female graduates demonstrate significantly higher sensitivity to the cost of living. These findings provide a high-resolution lens on the global ``brain circulation,'' highlighting the destination-specific comparative advantages that govern high-skilled relocation.
 
\end{abstract}

\maketitle

\section{Introduction}

The mobility of high-skilled professionals has emerged as an increasingly large and critical component of international migration \cite{Verbik2007InternationalSM}, with the population of highly educated migrants growing three times faster than less-educated groups between 1990 and 2010 \cite{PekkalaKerr2016GlobalTF}. By 2020, the majority of immigrants to Organization for Economic Cooperation and Development (OECD) countries possessed a tertiary education, as the rate of skilled emigration continues to exceed overall emigration rates due to the heightened human and financial capital associated with elite mobility \cite{dAiglepierre2020AGP, 2022GlobalMG}. This professional migration represents a vital labor source for countries facing advanced skills shortages and shrinking native-born workforces, directly influencing economic development, innovation, and the global circulation of knowledge \cite{DocquierRapoport2012, Garca2019ReseaHM}. Central to this phenomenon is international student mobility, which has nearly tripled over the last two decades. While the number of mobile tertiary students reached 2.2 million in 2001 \cite{Verbik2007InternationalSM}, it climbed to over 6.39 million by 2021—defying pandemic-era expectations with continued growth.\footnote{International Organization for Migration (IOM), \textit{World Migration Report 2024}, Chapter 2: International Students. Available at: \url{https://worldmigrationreport.iom.int/what-we-do/world-migration-report-2024-chapter-2/international-students}. Accessed 8 February 2026.} This population remains notably gendered; although the gap has narrowed, male students still comprise 52\% of the global total compared to 47\% for females.\footnote{In 2001, there were around 1 million internationally mobile female students (45\% of the total) and 1.2 million male students (54\%). While this gap has narrowed over the last 20 years, the number of internationally mobile female students remains lower than that of male students; in 2021, around 3 million students were female (47\%) and males comprised around 3.4 million (52\%). Source: UNESCO Institute for Statistics. Accessed 8 February 2026}
Historically, 70\% of these students have been concentrated in the US, UK, Germany, France, and Australia, with the US remaining the primary destination hosting over 833,000 students in 2021 and seeing an additional 8\% growth in the 2023/24 academic year.\footnote{International student totals grew by 8\% in the 2023/24 academic year, building on 12\% growth in 2022/23 and 4\% in 2021/22. This includes a 7\% increase in graduate enrollment and a 17\% surge in students pursuing employment via Optional Practical Training (OPT). See IIE (2023) \textit{Fall 2023 Snapshot on International Student Enrollment} Available at: \url{https://www.iie.org/wp-content/uploads/2023/11/Fall-2023-Snapshot.pdf}. Accessed 8 February 2026.}
However, as the market undergoes a structural shift toward emerging hubs in Asia and the Middle East, the strategic need for Western economies to retain international talent via pathways like Optional Practical Training (OPT) has intensified. 

Despite the wealth of literature on brain drain \cite{Docquier2007BrainDI}, brain
gain \cite{Kapur2005GiveUY} and brain circulation \cite{Wiesel2014FellowshipsTB,  Appelt2015WhichFI}, there remains a critical lack of reliable data regarding the specific trajectories of students from diverse origins \cite{Akbaritabar2023GlobalFA, Sanliturk2023GlobalPO}.
Despite the increasing scale of these flows, many dynamics related to skilled migration remain poorly understood, primarily due to persistent data limitations in administrative sources that lack granular disaggregation by gender and skill level \cite{Kofman2014TowardsAG, Donato2014TheDD, Lutz2014WorldPA}. High-quality migration and mobility data are notoriously difficult to obtain \cite{Skeldon2006InterlinkagesBI}; even widespread harmonization efforts such as the Integrated Public Use Microdata Series (IPUMS) \cite{Alexander2020CombiningSM} often reveal paradoxical records between neighboring nations due to divergent registration and digitization practices. 

Traditionally, researchers have attempted to remedy these gaps by treating migration as a residual ``error term'' calculated from the difference between population growth and vital statistics \cite{CanudasRomo2022TheCO}. However, such estimates fail to measure actual migration events and are prone to significant inaccuracies, particularly regarding the non-linear, intermediary steps taken by high-skilled cohorts \cite{Drouhot2022ComputationalAT}. To address this scarcity of reliable longitudinal data, recent studies have turned to digital trace data, high-resolution signals generated by social media and professional networks \cite{Zagheni2017LeveragingFA, CoimbraVieiraEtAl2022, Latour2007DigitalTraces}. Among these, the LinkedIn Advertising platform emerges as a uniquely robust source for studying professional migration, as it captures information on educational backgrounds, employment changes, and relocation behavior for over one billion users \cite{Kashyap2021HasDW, Perrotta2022OpennessTM}. Previous research has successfully leveraged LinkedIn to estimate the spatial distribution of university graduates \cite{Heo2023InvestigatingTD}, monitor the labor market integration of skilled migrants \cite{Ganguli2020TheRO}, and document international ``brain circulation'' through professional connectivity \cite{State2014MigrationOP, Kalhor2023GenderGI, Jacobs2024GlobalGG}. 
Crucially, it has been validated the platform's ability to mirror ``ground truth'' gender gaps reported by the International Labour Organization (ILO) \cite{Kashyap2021AnalysingGP}. 

By leveraging these digital traces from LinkedIn—the world’s largest professional networking platform with over 900 million members in over 200 countries,\footnote{https://learning.linkedin.com/content/dam/me/business/en-us/amp/learning-solutions/images/wlr-2024/LinkedIn-Workplace-Learning-Report-2024.pdf, Accessed: 8 February, 2026} we access novel data via the Advertising and Recruiter platforms. These sources provide timely, aggregate-level information about professional migrants at a significantly richer level of detail than traditional survey data \cite{Smith2020MillennialsUA}.\footnote{LinkedIn, ``Statistics \& Facts,'' \url{https://www.statista.com/topics/951/linkedin/} (accessed February 9, 2026).} 
In this study, we address existing research gaps through four primary objectives: mapping the mobility of students from QS World Ranked universities\footnote{\url{https://www.topuniversities.com/university-rankings}, Accessed: 8 February 2026} 
across 102 countries (RQ1); identifying demographic variances in ``who moves'' disaggregated by gender and age (RQ2); tracking the selectivity and distribution of these migrants across specific industries (RQ3); and finally, analyzing the socio-economic and cultural drivers that influence these global flows (RQ4).

\section*{Results and Discussion}
\subsection*{RQ1: Global topology of student mobility across 102 countries}
In this study, we define High-Potential Mobility as users who attended QS World University Rankings institutions and subsequently relocated internationally (LinkedIn’s “current address” measure, interpreted as an estimate of recently arrived professional migrant stock in destination countries) \cite{worldbank_linkedin_2018_methodology}. 
We compile the QS list of top-ranked universities (QS World University Rankings 2026: 1,504 institutions across 107 locations\footnote{QS Quacquarelli Symonds (QS), ``QS World University Rankings 2026: Top global universities,'' \textit{Top Universities}, accessed \textit{[DATE]}, \url{https://www.topuniversities.com/world-university-rankings}. The 2026 rankings include over 1,500 universities across more than 100 locations worldwide.}). Countries not available in the platform query interface (e.g., Russia, Iran, Syria, Northern Cyprus) and universities from these countries are excluded from the country-level coverage in our data collection. 

Data is collected as anonymous, aggregate-level estimates via the LinkedIn advertising with Campaign Manager platforms\footnote{https://www.linkedin.com/campaignmanager , Accessed: 9 Feb 2026}. To protect individual identifiers, the platform provides information only above a threshold of 300 users; targeted queries below this threshold do not return results. There are acknowledged trade-offs between the level of detail and data sparsity in aggregate-level data \cite{Kashyap2021HasDW}, particularly for specific industry categories and countries with smaller population sizes.
LinkedIn estimates user age based on profile information
such as years since college graduation and years in the labor force. We construct three age groups for our analysis: 18–24; 25–34;
35–55 and 55+ years old.  LinkedIn does not assign a gender to all users; on average, 75 percent of LinkedIn users are labeled as male
or female on the Advertising platform. We analyze the two gender categories provided by LinkedIn (male and female) but recognize the need for
future research on nonbinary users if this measure becomes available \cite{Lagos2022HasTB}. Despite these limitations, the data shows patterns consistent with analysis conducted by the World Bank on talent migration trends and industry composition on LinkedIn \cite{Zhu2018DataI, worldbank_linkedin_2018_methodology}, increasing our confidence in the present dataset. Relocated members are identified based on a country change in the location of employment between two job titles, which we interpret as an estimate of the recently arrived professional migrant stock in destination countries \cite{Jacobs2024GlobalGG, Vieira2025ForcedMA}.

\begin{figure*}[ht]
    \centering
    \includegraphics[width=1\linewidth]{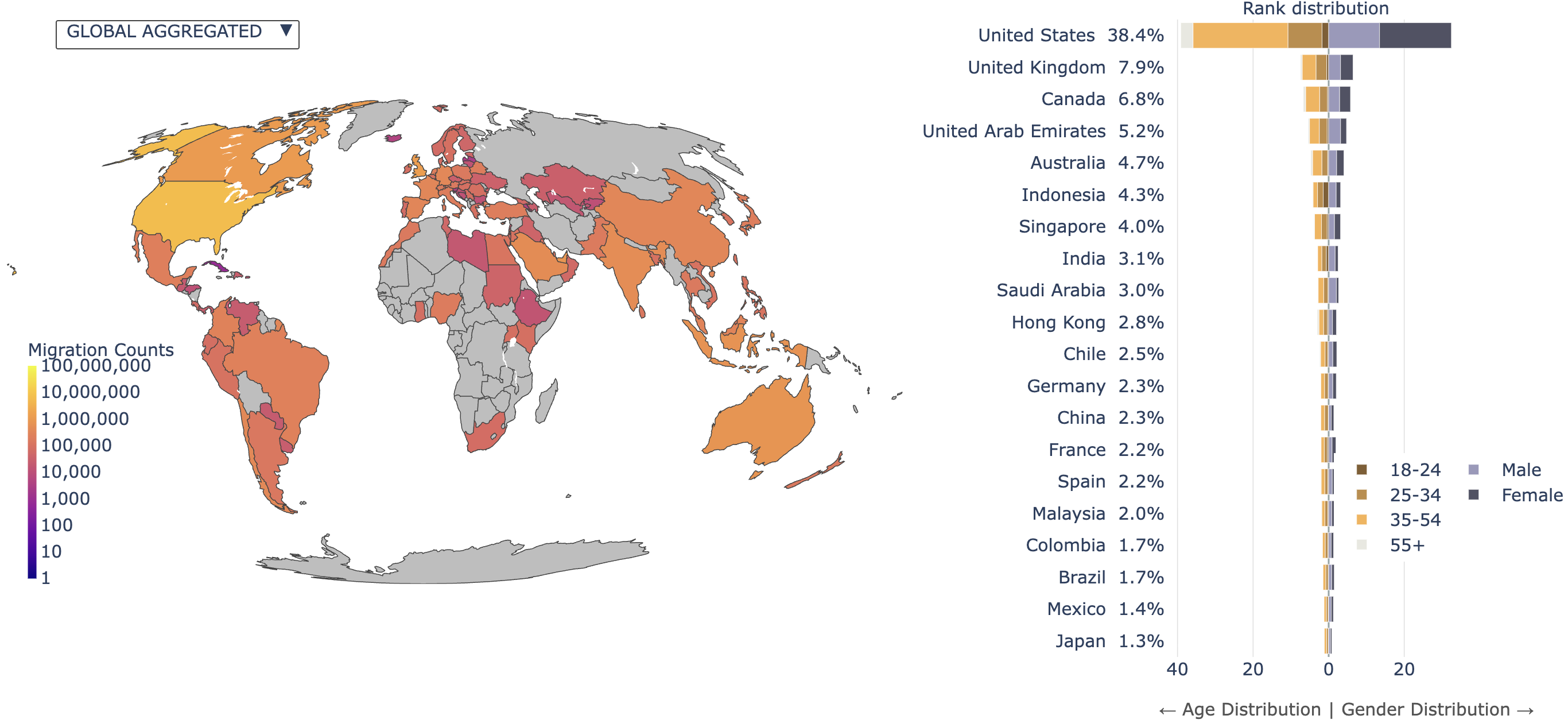}
    \caption{Global Topology and Demographic Composition of High-Potential Mobility.
The map illustrates the global distribution of internationally mobile graduates from QS-ranked universities, with destination countries colored by total student mobility counts. The accompanying Rank Distribution (right) highlights the top 20 destination countries by global share. Each bar in the rank distribution is disaggregated by Age Distribution (left-facing: 18–24, 25–34, 35–54, and 55+ cohorts) and Gender Distribution (right-facing: platform-inferred male and female shares), revealing a mobility pool predominantly concentrated in prime working ages (25–54) and showing varying degrees of gender balance across major destination hubs (see \href{https://tabiatanzin.github.io/migration-dashboard/}{interactive dashboard}).}
    \label{fig:aggregated_map}
\end{figure*}

The destination preferences of elite graduates reveal a highly concentrated topology of global talent flows, as shown in Figure \ref{fig:aggregated_map}, which displays a global density map (left) and a detailed rank distribution (right). The United States acts as the primary hub, capturing $38.4\%$ of the mobile elite. This dominance is driven by the 35–54 age bracket ($25.11\%$), suggesting the U.S. is a definitive settlement hub for established professionals. Traditional OECD nations follow, including the UK ($7.9\%$), Canada ($6.8\%$), and Australia ($4.7\%$). Notably, these Western hubs demonstrate robust gender integration; in the U.S., the female share ($19.01\%$) outpaces the male share ($13.47\%$), reflecting an ecosystem that successfully retains high-potential female talent.A shifting landscape is evidenced by powerful regional magnets diversifying the network. The UAE has emerged as a major node with a $5.2\%$ share, followed by Singapore ($4.0\%$), India ($3.1\%$), and Saudi Arabia ($3.0\%$). As shown in the Rank Distribution in Figure \ref{fig:aggregated_map}, these hubs display unique demographic profiles; while the UAE and Saudi Arabia show a male skew ($3.15\%$ and $2.04\%$ respectively), Indonesia ($4.3\%$) attracts a much younger cohort. Its 18–24 ($1.44\%$) and 25–34 ($1.51\%$) groups represent a higher proportion of intake than Western counterparts, signaling that young talent is increasingly drawn to high-growth regional powerhouses.The inclusion of Chile ($2.5\%$), Malaysia ($2.0\%$), and Colombia ($1.7\%$) in the top 20 highlights a transition toward a multi-polar network. These "middle-power'' nodes are successfully competing for early-career professionals, with Chile and Malaysia maintaining stable captures of the 25–34 age cohort. This trend suggests that historical "brain drain'' is evolving into "brain circulation,'' where elite graduates navigate a globalized market where the Global South is becoming increasingly competitive, ultimately reshaping the future of global human capital distribution.

The interactive visualization enables a granular decomposition of talent mobility from specific origin country (See Supplementary File \ref{sec:origin}), revealing distinct migratory “fingerprints.” Outward mobility from the United States shows a pivot toward emerging markets (See Figure~\ref{fig:USA}): Indonesia captures $16.7\%$ of the U.S. top-20 destination share and is strongly youth-skewed, with the 18–24 and 25–34 cohorts together accounting for $10.79\%$, far exceeding the 55+ share ($0.67\%$). In contrast, flows to Chile ($12.2\%$) are more senior-heavy, with the 35–54 and 55+ cohorts comprising over half of Chile’s share, while other prominent U.S.-origin destinations such as Cyprus ($7.4\%$) and Brunei Darussalam ($6.0\%$) emphasize mid-career movers. Similarly, United Kingdom reveals a structure shaped by historical and economic ties (See Figure~\ref{fig:uk}): the United States is the dominant recipient of UK-based elite talent, capturing $28.9\%$ of the UK top-20 distribution, with an overwhelmingly professional profile (a combined $18.52\%$ aged 35+), while the UK also serves as a feeder to Commonwealth and regional markets including India ($7.5\%$), Australia ($6.8\%$), and Singapore ($6.1\%$). Although flows to Australia remain relatively gender-balanced, mobility to the United Arab Emirates ($6.4\%$) is distinctly male-skewed ($3.56\%$ male vs. $1.81\%$ female), consistent with broader regional patterns. Overall, the country-by-country breakdown highlights increasingly complex “brain circulation” dynamics: for UK-origin mobility, Hong Kong ($5.0\%$) and China ($3.9\%$ emerge as key nodes for the 25–34 cohort, while Malaysia ($5.0\%$) and Nigeria ($2.9\%$ point to strengthening North–South corridors, illustrating how destinations compete for specific segments of the global elite workforce by age, gender, and career stage.

\begin{figure}
    \centering
    \includegraphics[width=1\linewidth]{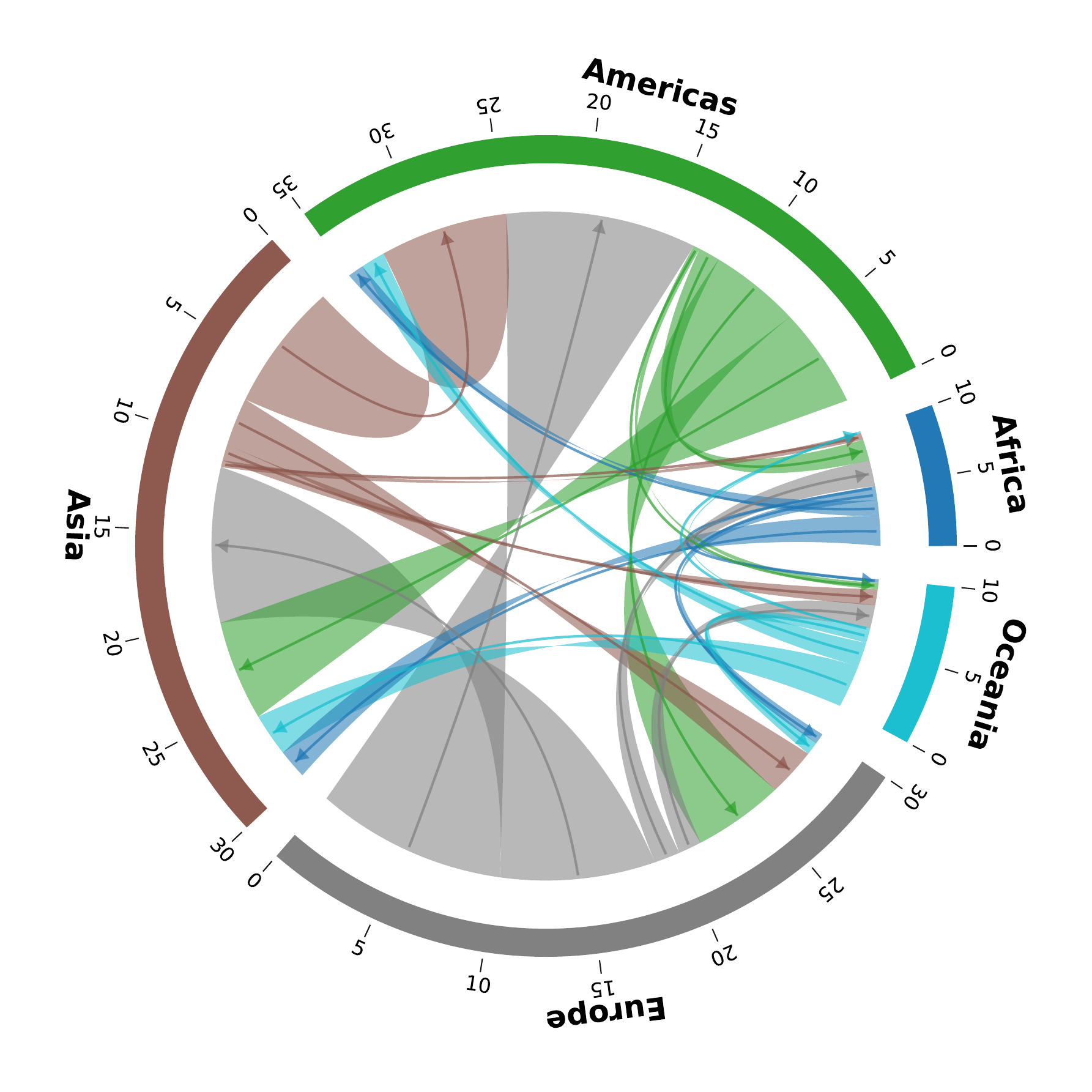}
    \caption{Directed Inter-Regional Flows of internationally mobile graduates from QS-ranked universities across five world regions. Each ribbon represents a directed flow, and its width at the origin indicates the share of an origin region’s total outflow that relocates to a given destination region.}
    \label{fig:regional_mapping}
\end{figure}

Demographic composition results (Supplementary Figure \ref{fig:supp_composition}) indicate strong imbalances by both gender and age. The gender distribution shows a small observed male share (2.55\%) in the platform-inferred aggregate (Supplementary Figure \ref{fig:supp_composition}). The age profile is concentrated in prime working ages: the 35–54 group comprises the largest share (23.10\%), followed by 25–34 (16.71\%) and 18–24 (2.45\%), while 55+ is consistently the smallest category (Supplementary Figure \ref{fig:supp_composition}). The region-level bar plot shows that total emigrant shares are unevenly distributed across regions and highlights stepwise gaps in ordered regional shares—most notably Oceania (↑16.68 percentage points), Asia (↑11.41), and Europe (↑9.37) (Supplementary Figure: Region-wise percent of total emigrants). Taken together, these composition patterns motivate examining where high-potential migrants relocate across the global system. To interpret the directionality of high-potential mobility, Figure \ref{fig:regional_mapping} summarizes directed inter-regional flows in percentage terms, where each ribbon represents the share of an origin region’s total outflow that relocates to a given destination (outgoing shares sum to 100\% within each origin). The resulting topology reveals clear destination preferences and asymmetric exchange patterns across regions. Asia’s outward trajectory is the most concentrated: 64.3\% of Asia’s outflow is directed to the Americas, followed by Europe (24.2\%) and Oceania (8.0\%), with a smaller share to Africa (3.5\%). European circulation exhibits a dual pull, with graduates splitting primarily between the Americas (48.3\%) and Asia (39.3\%), and smaller shares flowing to Africa (6.6\%) and Oceania (5.9\%). The Americas show a comparatively balanced pattern, sending most outflow to Asia (45.8\%) and Europe (41.4\%), with smaller fractions to Africa (9.8\%) and Oceania (3.0\%). Africa’s outflow is directed predominantly toward Asia (49.5\%), followed by the Americas (26.9\%), Europe (17.9\%), and Oceania (5.8\%). Finally, Oceania sends over half of its outflow to Asia (52.3\%), with additional flows to the Americas (29.5\%), Europe (15.9\%), and Africa (2.3\%), reinforcing Asia’s central role as a destination region across multiple origin-specific distributions.

\subsection*{RQ2:Gender Heterogeneity in Student Mobility}

\begin{figure*}[ht]
    \centering
    \includegraphics[width=1\linewidth]{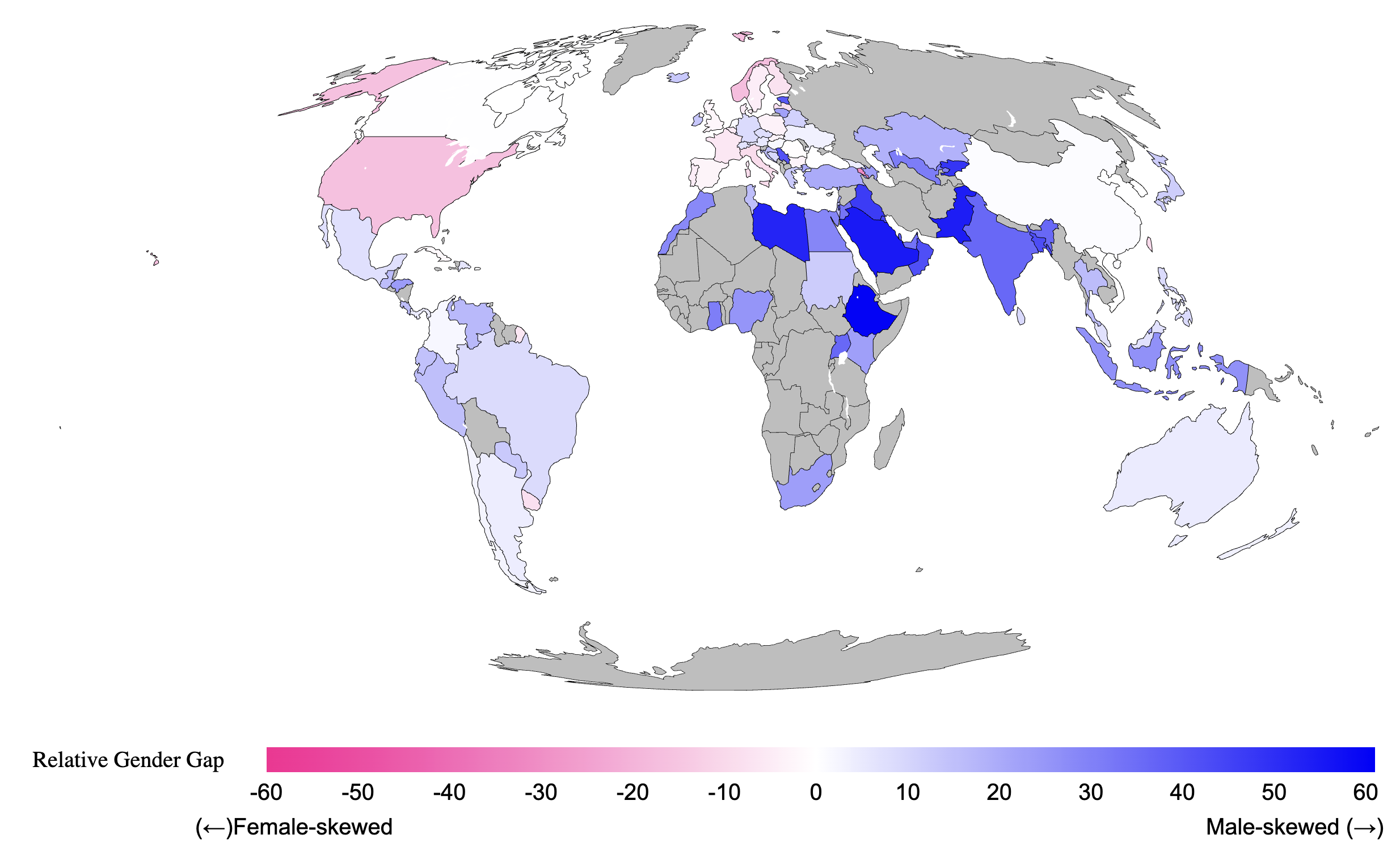}
    \caption{Global Relative Gender Gap (RGG) in High-Potential Mobility to show the gender imbalances among internationally mobile graduates from elite universities across 102 countries. The RGG score identifies destinations with male overrepresentation (blue, e.g., Ethiopia at +60.34\%) versus female overrepresentation (pink, e.g., Armenia at $-$30.77\%), while white areas signify near-parity (e.g., Canada at +0.22\%). (see \href{https://tabiatanzin.github.io/migration-dashboard/}{interactive dashboard})}
    \label{fig:gender}
\end{figure*}

We quantify gender imbalance in high-potential mobility using a Relative Gender Gap (RGG) score, where positive values indicate male overrepresentation and negative values indicate female overrepresentation among internationally mobile graduates from QS-ranked universities.\footnote{An RGG of $+X$\% indicates that men are overrepresented by $X$ percentage points relative to women in the observed mobility pool, whereas an RGG of $-X$\% indicates that women are overrepresented by $X$ percentage points.} At the global level, we estimate $\mathrm{RGG}=+3.16\%$, indicating a modest overall male advantage in international relocation among graduates of top-ranked universities\cite{UN_DESA_PopDiv_2015a, UN_DESA_PopDiv_2015b}. Despite near-parity globally, regional aggregates reveal pronounced divergence, consistent with variation in labor-market structure and institutional context. Destinations in Africa ($\mathrm{RGG}=+27.10\%$) and Asia ($\mathrm{RGG}=+23.10\%$) exhibit the largest male overrepresentation, indicating strongly male-skewed mobility into these regions. This pattern is consistent with broader evidence that migrant stocks in parts of the Middle East and North Africa (MENA) are heavily male---roughly two-thirds of international migrants in the region are male (2015), with especially high male shares in several Gulf states (e.g., 66\% male in Kuwait and 84\% male in Qatar) \cite{IOM_MENA_2016_DataSnapshot}.
while women comprise at least half of emigrants in other settings (e.g., the Occupied Palestinian Territories, Lebanon, and Jordan), underscoring how destination-specific labor markets and student mobility regimes shape gender composition. In contrast, the Americas display the most female-skewed profile ($\mathrm{RGG}=-10.35\%$), implying female overrepresentation among internationally mobile QS graduates. Europe ($\mathrm{RGG}=+1.10\%$) and Oceania ($\mathrm{RGG}=+4.40\%$) are closest to parity, with only modest male overrepresentation.

Consistent with these regional differences, Figure \ref{fig:gender} shows wide dispersion in gender balance across the 102 destination countries analyzed. The most male-skewed destinations include Ethiopia (+60.34\%), Saudi Arabia (+54.91\%), Pakistan (+54.13\%), Libya (+52.41\%), and Kyrgyzstan (+48.24\%); for example, RGG = +60.34\% indicates an extremely male-dominated mobility pool. Such extreme male skews are plausibly shaped by a combination of (i) gendered constraints on education and later mobility and (ii) sectoral labor demand that disproportionately recruits men. Evidence from Ethiopia, for example, documents persistent gender disadvantages for girls and young women—linked to entrenched gender norms and life-course constraints—which can reduce women’s ability to translate education into cross-border mobility relative to men.\footnote{Bekele Tefera and Paola Pereznieto, with Guday Emirie, \textit{Transforming the lives of girls and young women: Case study: Ethiopia} (London: Overseas Development Institute, August 2013). Available at: \texttt{https://odi.cdn.ngo/media/documents/8820.pdf}.}
Libya provides a clear illustration of how destination labor markets can generate male-heavy migrant pools: the IOM Displacement Tracking Matrix (DTM) reports that around 80\% of migrants in Libya are male, and that employed male migrants are concentrated in sectors including construction (reported as the largest category), alongside manufacturing and agricultural labor. For Kyrgyzstan, external evidence suggests an additional mechanism relevant to platform-based measurement: the World Bank reports that many female migrants from the Kyrgyz Republic are engaged in low-skilled domestic work abroad \cite{Parpieva2025TransformationOE}, which may reduce visibility in professional online platforms even when migrants have substantial human capital. This type of occupational downgrading or “skills mismatch” can plausibly contribute to male-skewed professional-platform mobility signals even when women migrate in large numbers.

At the same time, a substantial subset of countries cluster near parity, including Viet Nam ($-0.02\%$), South Korea ($-0.11\%$), Canada ($+0.22\%$), Puerto Rico ($+0.25\%$), and Romania ($-0.50\%$), with only modest skews in settings such as China (Mainland) ($+0.73\%$), the Netherlands ($+1.55\%$), and Hungary ($+1.64\%$). This near-zero cluster indicates that large gender gaps are not inevitable; many destinations appear to operate as approximately gender-neutral endpoints for high-potential movers.

At the female-skewed end of the distribution, the lowest RGG values are observed in Armenia ($-30.77\%$), Norway ($-17.48\%$), and the United States ($-16.28\%$), indicating substantial female overrepresentation among high-potential migrants in these destinations. In the United States, this is consistent with the broader demographic context in which immigrant women constitute a slight majority of the foreign-born population (51.3\% in 2021)~\cite{AIC2021},
which can support female-majority cohorts depending on the education-to-work pathways captured by platform measures. Armenia's strong female skew is consistent with documented male labor outmigration (often to Russia) that leaves women disproportionately represented in some rural communities, creating gender-selective mobility systems~\cite{Eurasianet2011, TheJournal2011}.
Norway's female skew similarly aligns with official statistics showing a large and growing female immigrant population; Statistics Norway reports 428,100 immigrant women at the beginning of 2023, corresponding to 16\% of all resident women~\cite{SSB2024}.

Taken together, these results indicate that while the global gender gap in high-potential mobility is small, gendered selectivity is strongly destination-specific. Extreme male overrepresentation is concentrated in a subset of destinations and is consistent with contexts where education-to-mobility pipelines are more gender-stratified and where labor demand is concentrated in male-dominated sectors (as seen in MENA aggregate patterns and in Libya’s DTM profile). Conversely, female-skewed destinations may reflect a combination of baseline immigrant composition (e.g., the United States), gendered mobility systems (e.g., male labor outmigration from Armenia), and destination contexts where female immigrant populations are growing (e.g., Norway). Importantly, the large cluster of near-parity destinations underscores that gender imbalance is not a universal feature of high-potential mobility but varies systematically across destinations and the pathways that connect education to international relocation.\\

\textbf{RQ2: Age Heterogeneity in Student Mobility}
We evaluate the demographic composition of high-potential mobility using the Age Dissimilarity Score ($D_c$), which quantifies how strongly a destination country’s migrant age distribution deviates from the global benchmark age shares. Values near zero indicate a comparatively ``standardized'' age structure dominated by prime working ages, whereas higher values indicate age-selective inflows concentrated in specific cohorts (e.g., student-heavy or older-skewed profiles). Across all destinations, the mean global dissimilarity is $D_c = 0.1663$, indicating moderate but meaningful age-structure heterogeneity across the global mobility system. Regional aggregates show clear differences in demographic stability and, by implication, in labor-market integration potential. Europe exhibits the lowest regional dissimilarity ($D_c = 0.0772$), consistent with an age profile that closely tracks the global benchmark and reflects a predominantly working-age inflow across multiple corridors. The Americas show a higher deviation ($D_c = 0.1286$) but remain strongly concentrated in active ages; in the United States, 77.1\% of the foreign-born population is of working age (16--64), compared with 60.9\% among the U.S.-born population, implying a lower dependency burden among migrants and a greater potential fiscal contribution through labor-force participation and tax revenue~\cite{AIC2021_General}.
Asia ($D_c = 0.1307$) and Oceania ($D_c = 0.1200$) show similar levels of deviation, consistent with corridor-specific selection that emphasizes younger working-age movers; in Oceania, mobility is often concentrated in early-career cohorts (median age 37)~\cite{UN2013},
consistent with recruitment and post-study pathways that disproportionately attract younger adults~\cite{IOM_AsiaPacific}.
Africa ($D_c = 0.1198$) also departs from the benchmark, reflecting a comparatively youthful mobility profile; the median age of international migrants in Africa was 31 in 2017~\cite{UN2017},
consistent with migration systems shaped by younger cohorts and substantial intra-continental movement.

\begin{figure*}[ht]
    \centering
    \includegraphics[width=1\linewidth]{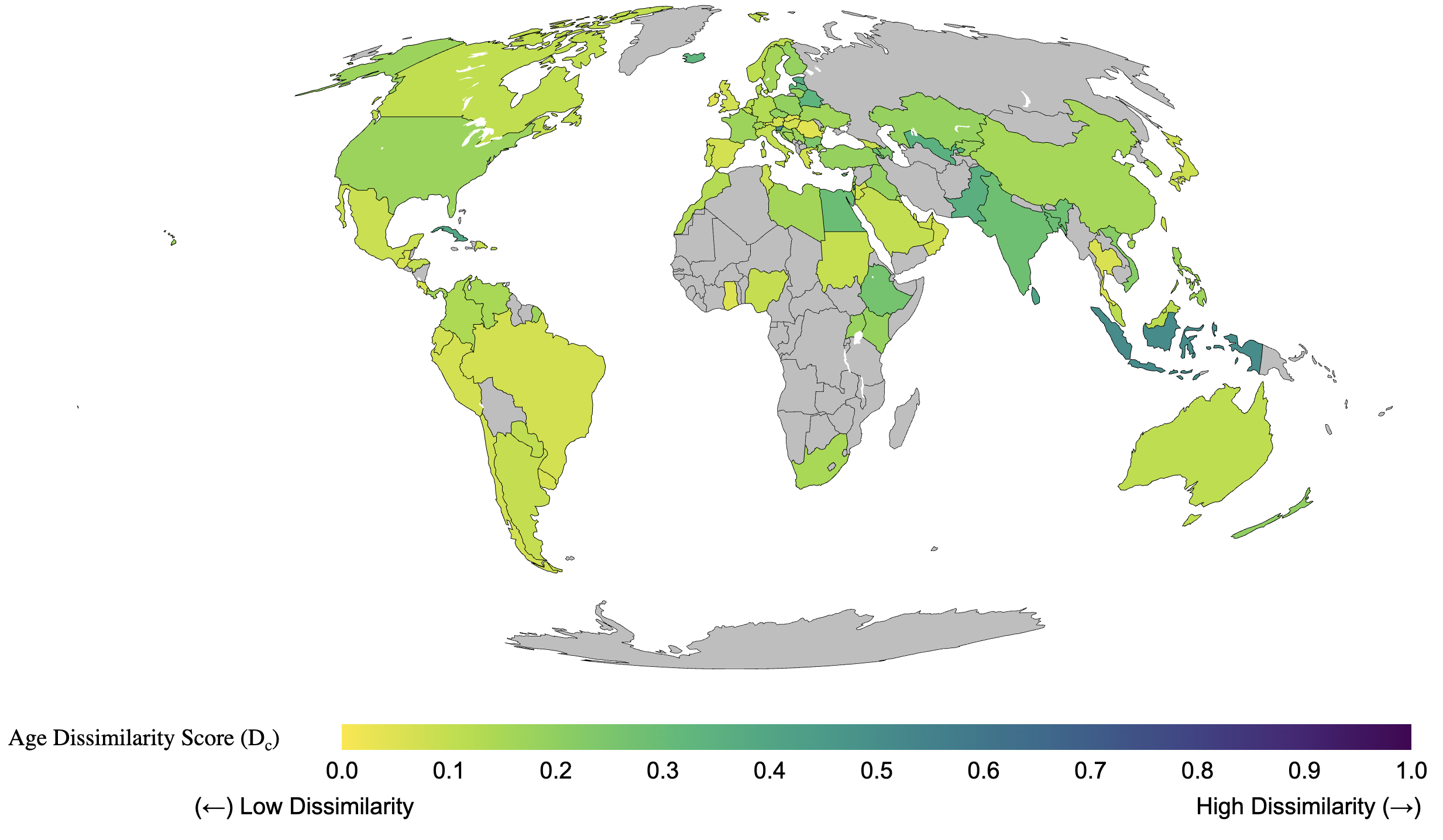}
    \caption{Global Age Imbalance ($D_c$ Score) of High-Potential Migrants to visualizes the Age Dissimilarity Score ($D_c$) across 102 destination countries, quantifying how strongly a nation's migrant age distribution deviates from the global benchmark. Destinations with low dissimilarity (yellow-green, e.g., Romania at $D_c=0.042$) reflect standardized working-age inflows, while high dissimilarity (dark teal/purple, e.g., Indonesia at $D_c=0.507$ and Slovenia at $D_c=0.505$) indicates age-selective mobility concentrated in specific cohorts due to labor recruitment regimes or historical corridors. (see \href{https://tabiatanzin.github.io/migration-dashboard/}{interactive dashboard})}
    \label{fig:age}
\end{figure*}

Figure \ref{fig:age} shows the cross-country differences in $D_c$, revealing outlier destinations where mobility is strongly cohort-concentrated, plausibly reflecting distinctive migration histories, labor demand, and crisis dynamics. Indonesia ($D_c = 0.507$) and Slovenia ($D_c = 0.505$) rank among the highest-discrepancy destinations, indicating pronounced deviation from the global benchmark. In Indonesia, age structure varies sharply by destination corridor: migrants residing in the Netherlands are substantially older, consistent with long-run historical and family-linked settlement channels, whereas migrants in Korea and Japan are heavily concentrated in younger cohorts, consistent with labor-market selection and contract-based recruitment that draws early-career movers \cite{OECD2022_IndonesianEmigrants}. Slovenia shows similarly strong age selectivity consistent with labor-linked entry, with migrants arriving at a relatively young average age (27 at immigration). Cuba ($D_c = 0.402$) represents another high-discrepancy case, consistent with crisis-driven, working-age--concentrated exits (``brain drain''): 77--80\% of emigrants fall between ages 15 and 59, a composition that plausibly accelerates demographic aging in the origin population by disproportionately removing economically active cohorts.

In contrast, several destinations exhibit low dissimilarity scores, indicating age distributions that closely resemble the global benchmark and reflect comparatively standardized working-age inflows. Romania ($D_c = 0.042$), Ireland ($D_c = 0.047$), and Slovakia ($D_c = 0.049$) fall in this low-imbalance range. In Ireland, 52\% of immigrants are aged 25--44, consistent with broad inflows centered on prime labor-market ages rather than a narrow cohort channel. In Slovakia, foreign residents are concentrated in the 30--34 range and show relatively high educational attainment (26.3\% holding university degrees versus 17.9\% among the native population), consistent with selective inflows dominated by early- to mid-career professionals rather than student-only or older-skewed pathways.

Overall, age-structure variation indicates that high-potential mobility is shaped by corridor-specific selection mechanisms rather than a single global demographic profile. Europe’s low $D_c$ suggests relatively benchmark-aligned inflows across multiple corridors, while the United States illustrates how a working-age--heavy migrant profile can reduce dependency burdens and support fiscal contributions. By contrast, Indonesia and Slovenia highlight how historically embedded corridors and labor recruitment regimes can generate strongly cohort-selective inflows, and Cuba illustrates how crisis dynamics can concentrate mobility into prime working ages with potential downstream consequences for origin-country aging. Because pathway composition can shift both age and gender distributions through sectoral sorting and mobility constraints, demographic patterns are most informative when interpreted jointly across $D_c$, RGG, and directed flow topology.\\

\textbf{RQ3: Professional Distribution of High-Skilled Migrant Graduates }

The professional integration of internationally mobile graduates exhibits a highly structured, domain-specific pattern across all 50 host economies, indicating that cross-border mobility is shaped less by uniform “global” demand than by destination-specific comparative advantage.

\begin{figure}[t!]
    \centering
    \includegraphics[width=1\linewidth]{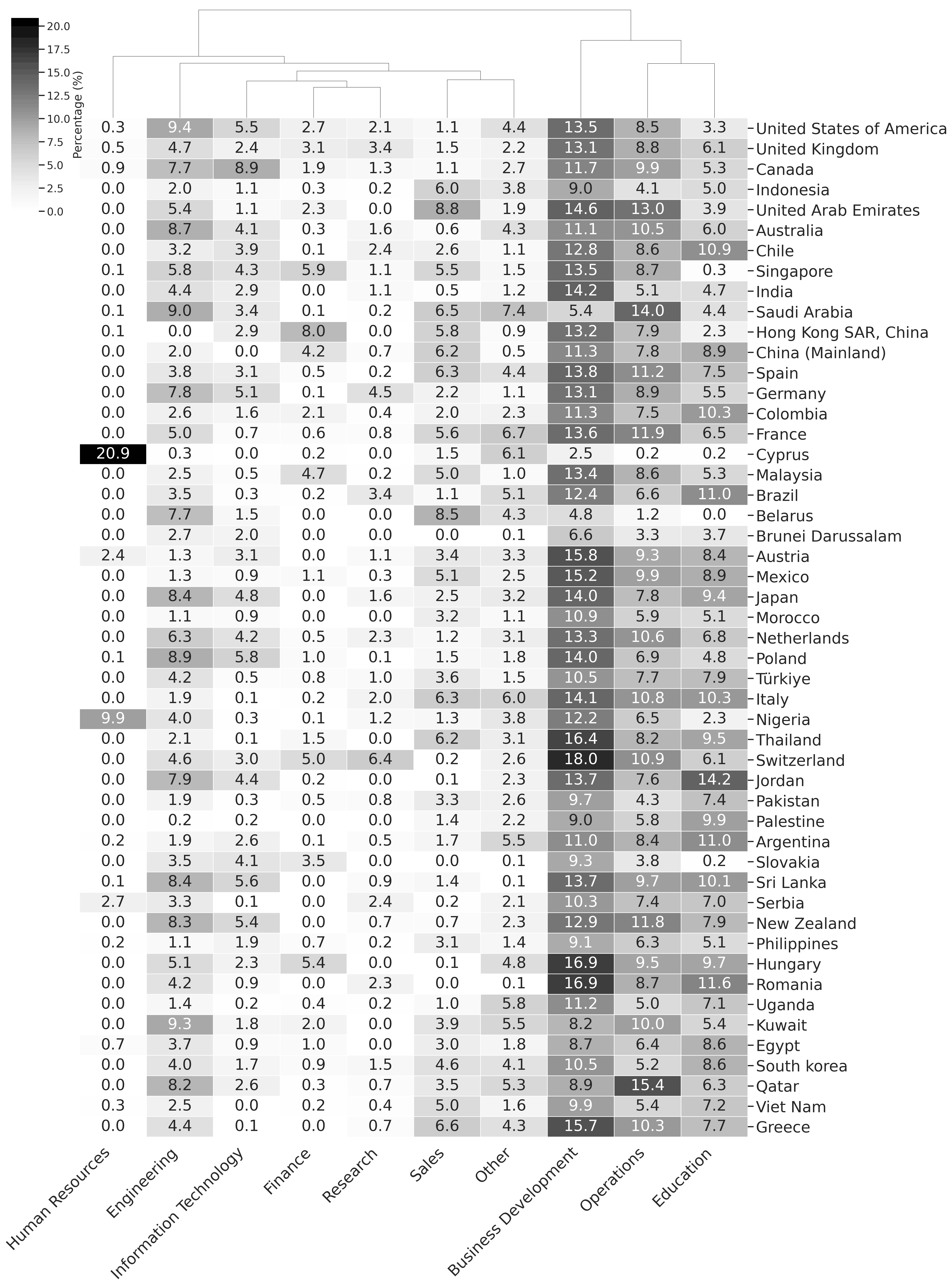}
    \caption{Professional Distribution of High-Skilled Migrant Graduates of the percentage of mobile graduates across 10 domains in 50 destination countries. Business Development and Operations are universal entry channels (e.g., Oman, Thailand $> 15\%$), while Engineering and IT are concentrated in innovation hubs like the U.S. (9.4\% Eng) and Canada (8.9\% IT).}
    \label{fig:prefession}
\end{figure}

Figure \ref{fig:prefession} shows across  nearly every country, Business Development and Operations function as the dominant entry channels for high-skilled talent, forming a commercial–operational engine that is particularly pronounced in service-scaling destinations such as the United Arab Emirates (14.6\% Business Development, 13.0\% Operations), Oman (16.9\% Business Development), and Thailand (16.4\% Business Development), and remains central even in secondary European markets like Austria (15.8\% Business Development), Greece (15.7\%), and Italy (14.1\%), where migrant absorption primarily reinforces commercial growth rather than deep technical specialization. In contrast, the pull for Engineering and Information Technology is sharply concentrated in innovation hubs that selectively absorb graduates into technical roles, including the United States (9.4\% Engineering, 5.5\% IT), Germany (7.8\% Engineering), Canada (8.9\% IT, 7.7\% Engineering), Israel (8.4\% Engineering, 5.6\% IT), as well as Japan (8.4\% Engineering) and Singapore (5.8\% Engineering), reflecting targeted demand tied to R\&D ecosystems and high-tech industry capacity. High-value knowledge-economy functions—especially Finance and Research—are concentrated in a smaller set of global nodes, most notably Hong Kong SAR (Finance 8.0\%), Switzerland (Finance 5.0\%, Research 6.4\%), and the United Kingdom (Research 3.4\%), while Education emerges as a major pathway in several Latin American and Asian destinations, including Jordan (14.2\%), Brazil (11.0\%), Argentina (11.0\%), Chile (10.9\%), China (Mainland) (8.9\%), and South Korea (8.6\%). Finally, the distributions reveal distinct niche outliers and emerging footprints, such as Cyprus as a strong anomaly with a concentrated focus on Human Resources (20.9\%), Poland as a secondary European technical magnet (IT 5.8\%), and developing host markets like Nigeria (Engineering 4.0\%, Education 2.3\%) and Uganda (Business Development 11.2\%, Education 7.1\%) where professional integration is still consolidating around foundational commercial and education sectors. Taken together, the results shows that the occupational composition of high-skilled graduate mobility is an adaptive response to destination-specific labor-market demand, with countries selectively integrating migrant talent into functions that mirror their national economic specialization.\\

\textbf{RQ4: Influencing Factors of Global Student Mobility}

To investigate the socioeconomic and demographic drivers of international student mobility, we estimated a multivariate Ordinary Least Squares (OLS) regression using destination-country predictors capturing economic capacity (GDP per capita, poverty, cost of living, economic freedom, remittances, unemployment), human capital and public services (education, life expectancy, healthcare expenditure), governance and social structure (democracy, gender gap, social welfare spending), and risk and stability (global peace, climate performance, women peace and security), alongside baseline geography and demography (area, population density).
Across 102 destination countries using data from
World Population Review \footnote{World Population Review, accessed 1 February 2026: \url{https://worldpopulationreview.com/}.}
, the model explains 33.6\% of the variance in global student mobility ($R^2=0.336$; See Appendix \ref{sec: Regressor} for details).

Figure \ref{fig:influecing_factors} shows student mobility is significantly positively associated with larger destination area, higher GDP per capita, more democratic governance, greater economic freedom, higher remittance intensity, stronger education conditions, and higher societal peace, consistent with a destination pull mechanism combining capacity and stability. In contrast, population density and unemployment exhibit statistically significant negative associations, indicating that congestion and weaker labor-market conditions reduce destination attractiveness. After conditioning on core economic and institutional factors, the gender gap index and healthcare expenditure show small but statistically significant negative associations, plausibly reflecting overlap among development indicators rather than direct deterrent effects. Several predictors, including social welfare spending, poverty, cost of living, life expectancy, climate performance, and women peace and security, do not reach conventional significance in this specification, suggesting limited marginal explanatory power beyond the core capacity and stability measures.\\

\begin{figure}[ht!]
    \centering
    \includegraphics[width=1\linewidth]{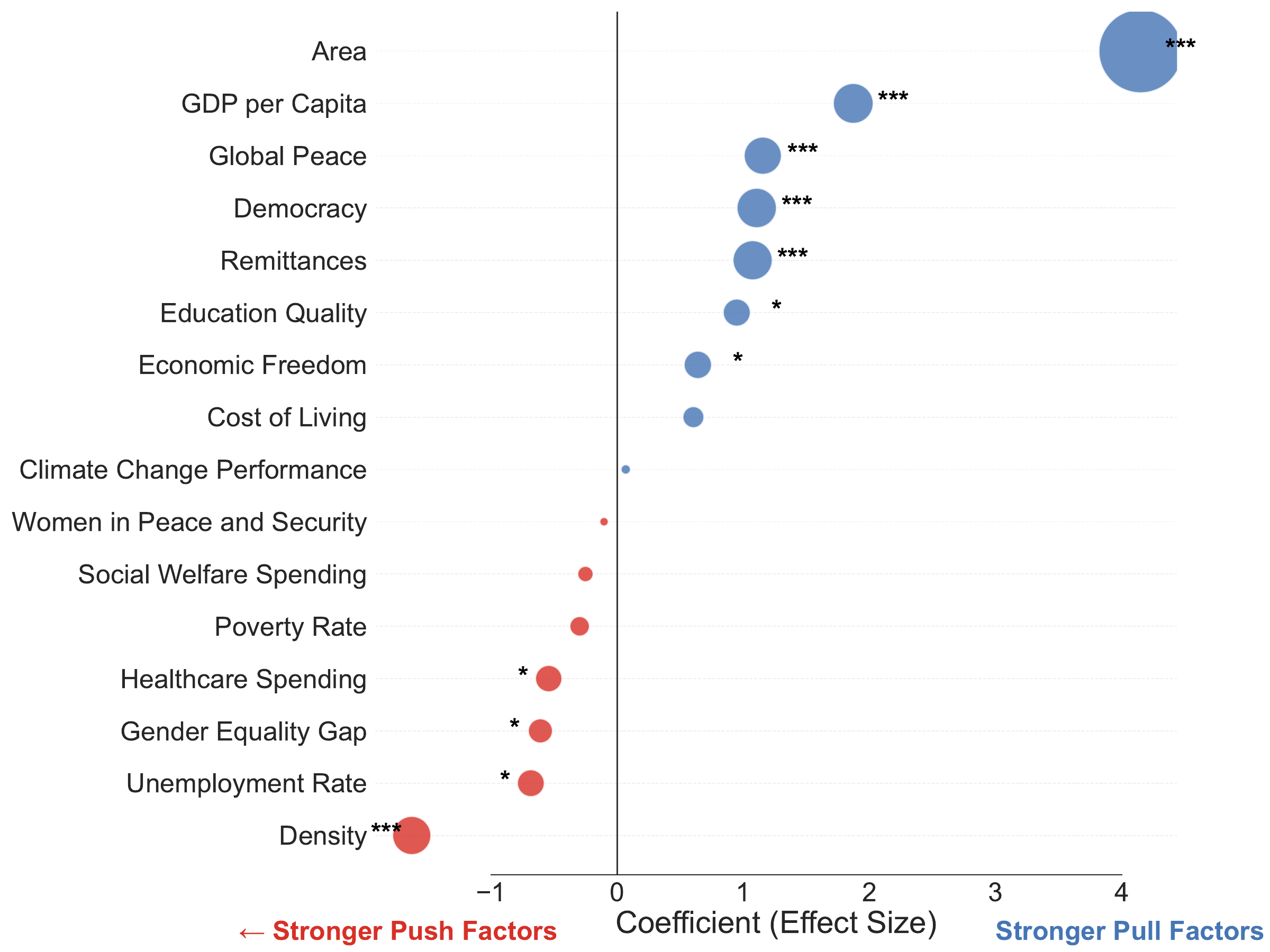}
    \caption{Drivers of Global High-Potential Mobility. Coefficient plot from a multivariate OLS regression ($R^2=0.336$) estimating factors associated with destination choices of internationally mobile elite graduates. Point size is proportional to the absolute \textit{t}-statistic, indicating the strength of statistical evidence for each predictor, and the horizontal axis reports coefficient estimates (effect sizes). Blue points denote positive associations (pull factors), such as geographic \textit{Area} and \textit{GDP per Capita}, while red points denote negative associations (deterrents), led by \textit{Population Density} and \textit{Unemployment Rate}. Significance levels: $^{***}p<0.001$, $^{*}p<0.05$.
}
    \label{fig:influecing_factors}
\end{figure}

To further examine heterogeneity in destination ``pull'' factors, we disaggregated mobility flows by gender and estimated models for male and female cohorts across 102 destination countries. Model performance is similar for both groups (male: $R^2=0.252$; female: $R^2=0.240$), indicating that the same macro-structural predictors explain comparable shares of variation in gender-specific mobility (See Appendix \ref{sec:gender_regression}). Figure~\ref{fig:gender_cofficient} summarizes coefficient concordance by plotting male against female estimates for each predictor. Most covariates cluster near the identity line, implying broadly symmetric associations across genders. In particular, destination area, GDP per capita, and democratic governance exhibit consistently positive associations, reflecting a shared preference for destinations with greater economic capacity and stronger institutions. Likewise, population density and unemployment are negatively associated with both male and female mobility, suggesting that structural crowding and weaker labor-market conditions reduce destination attractiveness for both groups.

Departures from the identity line reveal gender-differentiated sensitivities. Female mobility is more strongly associated with cost of living and education conditions, and shows a significant negative association with poverty, indicating greater selectivity with respect to destination socioeconomic context. In contrast, male mobility aligns more closely with economic freedom and global peace, consistent with a stronger response to market openness and broad stability signals in the male model. Taken together, the gender-disaggregated results suggest that destination choice is governed by a common ``capacity and stability'' mechanism, but the relative salience of specific socioeconomic constraints differs by gender in ways consistent with gendered opportunity structures and differential exposure to destination risk and resource environments.

\section*{Limitations and future directions}

While our approach leverages aggregated audience estimates to study high-skilled mobility at scale, several limitations warrant caution. The data are subject to self-selection, since LinkedIn users are not representative of national populations and skew toward highly educated, urban, and digitally connected groups. Measures also rely on self-reported profile attributes (e.g., schools attended, location), which may be incomplete or inconsistently reported across countries and demographic groups. In addition, audience counts and demographic labels are generated by proprietary, evolving platform systems, so definitions and taxonomies may shift over time in ways that complicate comparability and replication; relatedly, changes to the platform interface or endpoints can disrupt reproducibility. Finally, although we use only anonymized, aggregate outputs subject to a minimum audience threshold (300 users), ethical risks remain, including the possibility of group-level harms if aggregate statistics are misinterpreted or misused.

These constraints motivate several directions for future work. Substantively, we plan to examine gendered occupational sorting by linking gender-disaggregated mobility to destination professional profiles to identify where gaps are largest (e.g., Engineering vs.\ Education) and how these patterns vary by region and institutional context. We also aim to extend the analysis by degree level (bachelor’s, master’s, PhD) to test whether destination ``pull'' factors differ across stages of human-capital accumulation. Methodologically, future work should triangulate platform-based estimates with administrative and survey benchmarks where available, develop sensitivity analyses for platform penetration and measurement error, and document collection procedures to improve transparency and reproducibility.

\section*{Materials and Methods}

\subsection*{Data collection}

We construct a bilateral student-mobility dataset from aggregate audience estimates provided by LinkedIn’s Campaign Manager (Advertising) platform. Although LinkedIn does not allow targeting users by ``countries previously lived in,'' it supports targeting by \textit{Member school}, enabling queries for members who studied at a specific higher-education institution and currently reside in a specified country. We use this feature to approximate education-based origins: for each origin country, we compiled the 2026 QS-ranked universities\footnote{QS World University Rankings, accessed 2 February 2026: \url{https://www.topuniversities.com/world-university-rankings}.} and mapped each institution to its LinkedIn School entity (e.g., \texttt{urn:li:school:<ID>}) via the platform’s school search endpoint. Using these school identifiers, we queried audience counts for members who (i) studied at a QS-ranked institution in origin country $X$ and (ii) currently live in destination country $Y$, generating an origin--destination mobility matrix over 102 destination countries (each origin paired with the other 101 possible destinations). While the QS rankings cover 107 countries, LinkedIn location targeting was not available for Russia, Iran, Syria, and Northern Cyprus, limiting our destination coverage to 102 countries (see Appendix for the list \ref{sec:country}). All outputs are anonymous, aggregate estimates; LinkedIn suppresses audiences below a privacy threshold of 300 users, so some small flows are censored.

In addition to total flows, we collected stratified audience counts by gender (male/female) and age (18--24, 25--34, 35--54, 55+), and we extracted destination-specific occupational composition using LinkedIn’s job-function taxonomy. For each destination, occupational categories are reported as shares of the migrant stock (e.g., Business Development, Operations, Engineering, Information Technology, Education, Finance), allowing us to characterize how host labor markets absorb internationally mobile graduates.


\subsection*{Relative Gender Gap (RGG)}

We quantify gender imbalance in high-potential mobility using a \textit{Relative Gender Gap} (RGG) score. For each unit $i$ (e.g., destination country or destination region), let $M_i$ and $F_i$ denote the total number of male and female migrants, obtained by summing gender-disaggregated LinkedIn audience estimates across all relevant origin--destination pairs. We define
\begin{equation}
\label{eq:rgg}
\mathrm{RGG}_i \;=\; 100 \times \frac{M_i - F_i}{M_i + F_i},
\end{equation}
where $\mathrm{RGG}_i>0$ indicates percentage of male overrepresentation, $\mathrm{RGG}_i<0$ indicates percentage of female overrepresentation, and $\mathrm{RGG}_i=0$ indicates parity. Region-level scores are computed by first aggregating countries within region $r$, $M_r=\sum_{j\in r} M_j$ and $F_r=\sum_{j\in r} F_j$, and then applying Eq.~\ref{eq:rgg}. The global RGG is computed analogously using totals $M=\sum_i M_i$ and $F=\sum_i F_i$.

\subsection*{Age dissimilarity score}

We quantify how strongly a destination’s age composition of internationally mobile graduates deviates using an \textit{Age Dissimilarity} score. Let $g_k$ denote the global share of migrants in age group $k \in \{18$--$24,\,25$--$34,\,35$--$54,\,55+\}$, computed as
\begin{equation}
g_k=\frac{\sum_i N_{ik}}{\sum_i T_i},
\end{equation}
where $N_{ik}$ is the number of migrants in age group $k$ for unit $i$ and $T_i=\sum_k N_{ik}$ is total migrants in $i$. For each unit $i$, we compute within-unit age shares $s_{ik}=N_{ik}/T_i$ and define
\begin{equation}
\label{eq:ads}
D_i=\sqrt{\frac{1}{3}\sum_{k}\frac{(s_{ik}-g_k)^2}{g_k}}.
\end{equation}
Larger $D_i$ values indicate greater deviation from the global age profile. We compute region-level scores by first aggregating age counts within each region and then applying the same equation \ref{eq:ads}.

\subsection*{Estimating destination-level professional profiles}

We estimate destination-level professional profiles by combining LinkedIn audience counts with job-function shares. For each origin--destination pair, LinkedIn provides the total migrant audience $N_{od}$ and the percentage distribution across function categories $p_{odk}$ (e.g., Engineering, Education, Finance). We convert these percentages into implied category counts,
\begin{equation}
N_{odk} \;=\; N_{od}\times \frac{p_{odk}}{100},
\end{equation}
and then aggregate across origins to obtain destination totals $N_{dk}=\sum_o N_{odk}$. Finally, we report each destination’s professional composition as shares of its migrant stock,
\begin{equation}
s_{dk} \;=\; 100\times \frac{N_{dk}}{\sum_k N_{dk}}.
\end{equation}

\section*{Data Availability}

All student-mobility datasets and destination-country demographic indicators used in this study are provided in CSV format in the \texttt{data/} folder of the repository\footnote{https://github.com/TabiaTanzin/migration-dashboard}. All analysis code, including preprocessing pipelines, professional-profile construction, and the regression models reported in the paper, is available in the \texttt{code/} folder.

\acknowledgments
The authors express their gratitude for helpful conversations with Julia Witte Zimmerman, Ashley Fehr, Calla Beauregard, Alejandro Javier Ruiz Iglesias, Mikaela Irene Fudolig. The authors acknowledge financial support from the National Science Foundation under Award Nos. 2242829 (VT EPSCoR SOCKS) and 2419830 (SaTC), and the Vermont Advanced Computing Core (VACC) for computational support.
\clearpage
\bibliographystyle{plain}
\bibliography{references}

\begin{thebibliography}{10}

\bibitem{2022GlobalMG}
Global migration, gender, and health professional credentials.
\newblock 2022.

\bibitem{Eurasianet2011}
Gayane Abrahamyan and Justyna Mielnikiewicz.
\newblock Armenia: A {W}oman's {W}orld in {O}ne {M}ountain {V}illage.
\newblock Eurasianet, March 7 2011.
\newblock Accessed: 2026-03-23.

\bibitem{Akbaritabar2023GlobalFA}
Aliakbar Akbaritabar, Tom Theile, and Emilio Zagheni.
\newblock Global flows and rates of international migration of scholars.
\newblock 2023.

\bibitem{Alexander2020CombiningSM}
Monica~J. Alexander, Kivan Polimis, and Emilio Zagheni.
\newblock Combining social media and survey data to nowcast migrant stocks in the united states.
\newblock {\em Population Research and Policy Review}, 41:1 -- 28, 2020.

\bibitem{AIC2021_General}
{American Immigration Council}.
\newblock Immigrants in the {U}nited {S}tates, 2021.
\newblock Accessed: 2026-03-23.

\bibitem{AIC2021}
{American Immigration Council}.
\newblock A snapshot of immigrant women in the united states, 2021.
\newblock Accessed: 2026-03-23.

\bibitem{Appelt2015WhichFI}
Silvia Appelt, Brigitte van Beuzekom, Fernando Galindo-Rueda, Roberto de, and Pinho.
\newblock Which factors influence the international mobility of research scientists.
\newblock 2015.

\bibitem{TheJournal2011}
Aoife Barry.
\newblock Women left behind as $98\%$ of men leave {A}rmenian village.
\newblock The Journal, March 18 2011.

\bibitem{CanudasRomo2022TheCO}
Vladimir Canudas-Romo, Tianyu Shen, and Collin~F. Payne.
\newblock The components of change in population growth rates.
\newblock {\em Demography}, 2022.

\bibitem{CoimbraVieiraEtAl2022}
C.~Coimbra~Vieira, M.~Fatehkia, K.~Garimella, I.~G. Weber, and E.~Zagheni.
\newblock Using facebook and linkedin data to study international mobility.
\newblock In A.~A. Salah, E.~E. Korkmaz, and T.~Bircan, editors, {\em Data Science for Migration and Mobility}, volume 251 of {\em Proceedings of the British Academy}, pages 141--158. Oxford University Press, Oxford, 2022.

\bibitem{Docquier2007BrainDI}
Fr{\'e}d{\'e}ric Docquier, Olivier Lohest, and Abdeslam Marfouk.
\newblock Brain drain in developing countries.
\newblock {\em The World Bank Economic Review}, 21:193--218, 2007.

\bibitem{DocquierRapoport2012}
Fr{\'e}d{\'e}ric Docquier and Hillel Rapoport.
\newblock Globalization, brain drain, and development.
\newblock {\em Journal of Economic Literature}, 50(3):681--730, 2012.

\bibitem{Donato2014TheDD}
Katharine~M. Donato, Bhumika Piya, and Anna~W. Jacobs.
\newblock The double disadvantage reconsidered: Gender, immigration, marital status, and global labor force participation in the 21st century 1.
\newblock {\em International Migration Review}, 48:335 -- 376, 2014.

\bibitem{Drouhot2022ComputationalAT}
Lucas~G. Drouhot, Emanuel Deutschmann, Carolina~Viviana Zuccotti, and Emilio Zagheni.
\newblock Computational approaches to migration and integration research: promises and challenges.
\newblock {\em Journal of Ethnic and Migration Studies}, 49:389 -- 407, 2022.

\bibitem{dAiglepierre2020AGP}
Rohen d’Aiglepierre, Anda~Mariana David, Charlotte Levionnois, Gilles Spielvogel, Michele Tuccio, and Erik~R. Vickstrom.
\newblock A global profile of emigrants to oecd countries.
\newblock 2020.

\bibitem{Ganguli2020TheRO}
Ina Ganguli, Shulamit Kahn, and Megan~J. MacGarvie.
\newblock The roles of immigrants and foreign students in us science, innovation, and entrepreneurship.
\newblock 2020.

\bibitem{Garca2019ReseaHM}
Tel{\'e}sforo~Ram{\'i}rez Garc{\'i}a.
\newblock Rese{\~n}a: High-skilled migration. drivers and policies.
\newblock {\em Revista Latinoamericana de Poblaci{\'o}n}, 13:162--165, 2019.

\bibitem{Heo2023InvestigatingTD}
Nayoung Heo, Hsin-Chieh Chang, and Guy~J. Abel.
\newblock Investigating the distribution of university alumni populations within south korea and taiwan based on data from the linkedin advertising platform.
\newblock {\em Cities}, 2023.

\bibitem{IOM_AsiaPacific}
{International Organization for Migration (IOM)}.
\newblock Age and sex breakdown of international migrants.
\newblock Asia-Pacific Migration Data Hub, 2026.
\newblock Accessed: 2026-03-23.

\bibitem{IOM_MENA_2016_DataSnapshot}
{International Organization for Migration (IOM), Regional Office for the Middle East and North Africa}.
\newblock Migration to, from and in the middle east and north africa: Data snapshot.
\newblock Report, April 2016.
\newblock Prepared by IOM Regional Office for the Middle East and North Africa.

\bibitem{Jacobs2024GlobalGG}
Elizabeth~M. Jacobs, Tom Theile, Daniela Perrotta, Xinyi Zhao, Athina Anastasiadou, and Emilio Zagheni.
\newblock Global gender gaps in the international migration of professionals on linkedin.
\newblock {\em Population and Development Review}, 51, 2024.

\bibitem{Kalhor2023GenderGI}
Ghazal Kalhor, Hannah Gardner, Ingmar Weber, and Ridhi Kashyap.
\newblock Gender gaps in online social connectivity, promotion and relocation reports on linkedin.
\newblock {\em ArXiv}, abs/2308.13296, 2023.

\bibitem{Kapur2005GiveUY}
Devesh Kapur.
\newblock Give us your best and brightest: The global hunt for talent and its impact on the developing world.
\newblock 2005.

\bibitem{Kashyap2021HasDW}
Ridhi Kashyap.
\newblock Has demography witnessed a data revolution? promises and pitfalls of a changing data ecosystem.
\newblock {\em Population Studies}, 75:47 -- 75, 2021.

\bibitem{Kashyap2021AnalysingGP}
Ridhi Kashyap and Florianne C.~J. Verkroost.
\newblock Analysing global professional gender gaps using linkedin advertising data.
\newblock {\em EPJ Data Science}, 10, 2021.

\bibitem{PekkalaKerr2016GlobalTF}
Sari~Pekkala Kerr, William~R. Kerr, Caglar Ozden, and Christopher Parsons.
\newblock Global talent flows.
\newblock {\em International Trade eJournal}, 2016.

\bibitem{Kofman2014TowardsAG}
Eleonore Kofman.
\newblock Towards a gendered evaluation of (highly) skilled immigration policies in europe.
\newblock {\em International Migration}, 52:116--128, 2014.

\bibitem{Lagos2022HasTB}
Danya Lagos.
\newblock Has there been a transgender tipping point? gender identification differences in u.s. cohorts born between 1935 and 2001.
\newblock {\em American Journal of Sociology}, 128:94 -- 143, 2022.

\bibitem{Latour2007DigitalTraces}
Bruno Latour.
\newblock Beware, your imagination leaves digital traces.
\newblock {\em Times Higher Literary Supplement}, April 2007.
\newblock application/pdf.

\bibitem{Lutz2014WorldPA}
Wolfgang Lutz, William~P. Butz, and Samir Kc.
\newblock World population and human capital in the twenty-first century.
\newblock 2014.

\bibitem{Note1}
International Organization for Migration (IOM), \protect \textit {World Migration Report 2024}, Chapter 2: International Students. Available at: \protect \url {https://worldmigrationreport.iom.int/what-we-do/world-migration-report-2024-chapter-2/international-students}. Accessed 8 February 2026.

\bibitem{Note10}
Bekele Tefera and Paola Pereznieto, with Guday Emirie, \protect \textit {Transforming the lives of girls and young women: Case study: Ethiopia} (London: Overseas Development Institute, August 2013). Available at: \protect \texttt {https://odi.cdn.ngo/media/documents/8820.pdf}.

\bibitem{Note11}
World Population Review, accessed 1 February 2026: \protect \url {https://worldpopulationreview.com/}.

\bibitem{Note12}
QS World University Rankings, accessed 2 February 2026: \protect \url {https://www.topuniversities.com/world-university-rankings}.

\bibitem{Note13}
https://github.com/TabiaTanzin/migration-dashboard.

\bibitem{Note2}
In 2001, there were around 1 million internationally mobile female students (45\% of the total) and 1.2 million male students (54\%). While this gap has narrowed over the last 20 years, the number of internationally mobile female students remains lower than that of male students; in 2021, around 3 million students were female (47\%) and males comprised around 3.4 million (52\%). Source: UNESCO Institute for Statistics. Accessed 8 February 2026.

\bibitem{Note3}
International student totals grew by 8\% in the 2023/24 academic year, building on 12\% growth in 2022/23 and 4\% in 2021/22. This includes a 7\% increase in graduate enrollment and a 17\% surge in students pursuing employment via Optional Practical Training (OPT). See IIE (2023) \protect \textit {Fall 2023 Snapshot on International Student Enrollment} Available at: \protect \url {https://www.iie.org/wp-content/uploads/2023/11/Fall-2023-Snapshot.pdf}. Accessed 8 February 2026.

\bibitem{Note4}
https://learning.linkedin.com/content/dam/me/business/en-us/amp/learning-solutions/images/wlr-2024/LinkedIn-Workplace-Learning-Report-2024.pdf, Accessed: 8 February, 2026.

\bibitem{Note5}
LinkedIn, ``Statistics \& Facts,'' \protect \url {https://www.statista.com/topics/951/linkedin/} (accessed February 9, 2026).

\bibitem{Note6}
\protect \url {https://www.topuniversities.com/university-rankings}, Accessed: 8 February 2026.

\bibitem{Note7}
QS Quacquarelli Symonds (QS), ``QS World University Rankings 2026: Top global universities,'' \protect \textit {Top Universities}, accessed \protect \textit {[DATE]}, \protect \url {https://www.topuniversities.com/world-university-rankings}. The 2026 rankings include over 1,500 universities across more than 100 locations worldwide.

\bibitem{Note8}
https://www.linkedin.com/campaignmanager , Accessed: 9 Feb 2026.

\bibitem{Note9}
An RGG of $+X$\% indicates that men are overrepresented by $X$ percentage points relative to women in the observed mobility pool, whereas an RGG of $-X$\% indicates that women are overrepresented by $X$ percentage points.

\bibitem{OECD2022_IndonesianEmigrants}
{OECD}.
\newblock A review of indonesian emigrants, 2022.

\bibitem{Parpieva2025TransformationOE}
Nurzhamal Parpieva.
\newblock Transformation of employment in the agricultural sector of kyrgyzstan: Challenges and prospects.
\newblock {\em Bulletin of the Kyrgyz National Agrarian University}, 2025.

\bibitem{Perrotta2022OpennessTM}
Daniela Perrotta, Sarah~C. Johnson, Tom Theile, Andr{\'e} Grow, Helga de~Valk, and Emilio Zagheni.
\newblock Openness to migrate internationally for a job: Evidence from linkedin data in europe.
\newblock In {\em International Conference on Web and Social Media}, 2022.

\bibitem{Sanliturk2023GlobalPO}
Ebru Sanliturk, Emilio Zagheni, Maciej~Jan Dańko, Tom Theile, and Aliakbar Akbaritabar.
\newblock Global patterns of migration of scholars with economic development.
\newblock {\em Proceedings of the National Academy of Sciences of the United States of America}, 120, 2023.

\bibitem{Skeldon2006InterlinkagesBI}
Ronald Skeldon.
\newblock Interlinkages between internal and international migration and development in the asian region.
\newblock {\em Population Space and Place}, 12:15--30, 2006.

\bibitem{Smith2020MillennialsUA}
Stephanie~A. Smith and Brandi~A. Watkins.
\newblock Millennials’ uses and gratifications on linkedin: Implications for recruitment and retention.
\newblock {\em International Journal of Business Communication}, 60:560 -- 586, 2020.

\bibitem{State2014MigrationOP}
Bogdan State, Mario Rodr{\'i}guez, Dirk Helbing, and Emilio Zagheni.
\newblock Migration of professionals to the u.s. - evidence from linkedin data.
\newblock In {\em Social Informatics}, 2014.

\bibitem{SSB2024}
{Statistics Norway (SSB)}.
\newblock Immigrant women and their {N}orwegian-born daughters: {D}emography, education, labour and income.
\newblock Technical report, Statistisk sentralbyrå, February 19 2024.
\newblock Accessed: 2026-03-23.

\bibitem{UN2013}
{United Nations}.
\newblock International {M}igration 2013: {A}ge and {S}ex {D}istribution.
\newblock Technical Report 2015/4, Department of Economic and Social Affairs, Population Division, 2015.

\bibitem{UN2017}
{United Nations}.
\newblock International {M}igration {R}eport 2017: {H}ighlights.
\newblock Technical report, Department of Economic and Social Affairs, Population Division, New York, 2017.

\bibitem{UN_DESA_PopDiv_2015a}
{United Nations, Department of Economic and Social Affairs, Population Division}.
\newblock Trends in international migrant stock: The 2015 revision.
\newblock United Nations database, 2015.
\newblock POP/DB/MIG/Stock/Rev.2015.

\bibitem{UN_DESA_PopDiv_2015b}
{United Nations, Department of Economic and Social Affairs, Population Division}.
\newblock World population prospects: The 2015 revision.
\newblock DVD Edition, 2015.

\bibitem{Verbik2007InternationalSM}
Line Verbik and Veronica Lasanowski.
\newblock International student mobility: Patterns and trends.
\newblock 2007.

\bibitem{Vieira2025ForcedMA}
Carolina~Coimbra Vieira, Ebru Sanliturk, and Emilio Zagheni.
\newblock Forced migration and information-seeking behavior on wikipedia: Insights from the ukrainian refugee crisis.
\newblock {\em ArXiv}, abs/2512.01692, 2025.

\bibitem{Wiesel2014FellowshipsTB}
Torsten~N. Wiesel.
\newblock Fellowships: Turning brain drain into brain circulation.
\newblock {\em Nature}, 510:213 -- 214, 2014.

\bibitem{worldbank_linkedin_2018_methodology}
{World Bank Group} and {LinkedIn}.
\newblock World bank group--linkedin data insights: Jobs, skills, and migration trends—methodology and validation results, 2018.

\bibitem{Zagheni2017LeveragingFA}
Emilio Zagheni, Ingmar Weber, and Krishna~P. Gummadi.
\newblock Leveraging facebook's advertising platform to monitor stocks of migrants.
\newblock {\em Population and Development Review}, 43:721--734, 2017.

\bibitem{Zhu2018DataI}
Tingting Zhu, Alan Fritzler, and Jan Alexander~Kazimierz Orlowski.
\newblock Data insights : Jobs, skills and migration trends methodology and validation results.
\newblock 2018.

\end{thebibliography}

\clearpage

\appendix

\section{Student Mobility by Origin Country}
\label{sec:origin}
 The interactive mobility map shows origin-specific student mobility patterns for each country included in the dataset. For every origin country, we identify the top 20 destination countries based on relative student mobility share and decompose these flows by age cohort (18--24, 25--34, 35--54, 55+) and gender (platform-inferred male and female). The total share for each destination represents its proportion within the origin country’s top 20 outward mobility distribution.

\begin{figure*}
    \centering
    \includegraphics[width=1\linewidth]{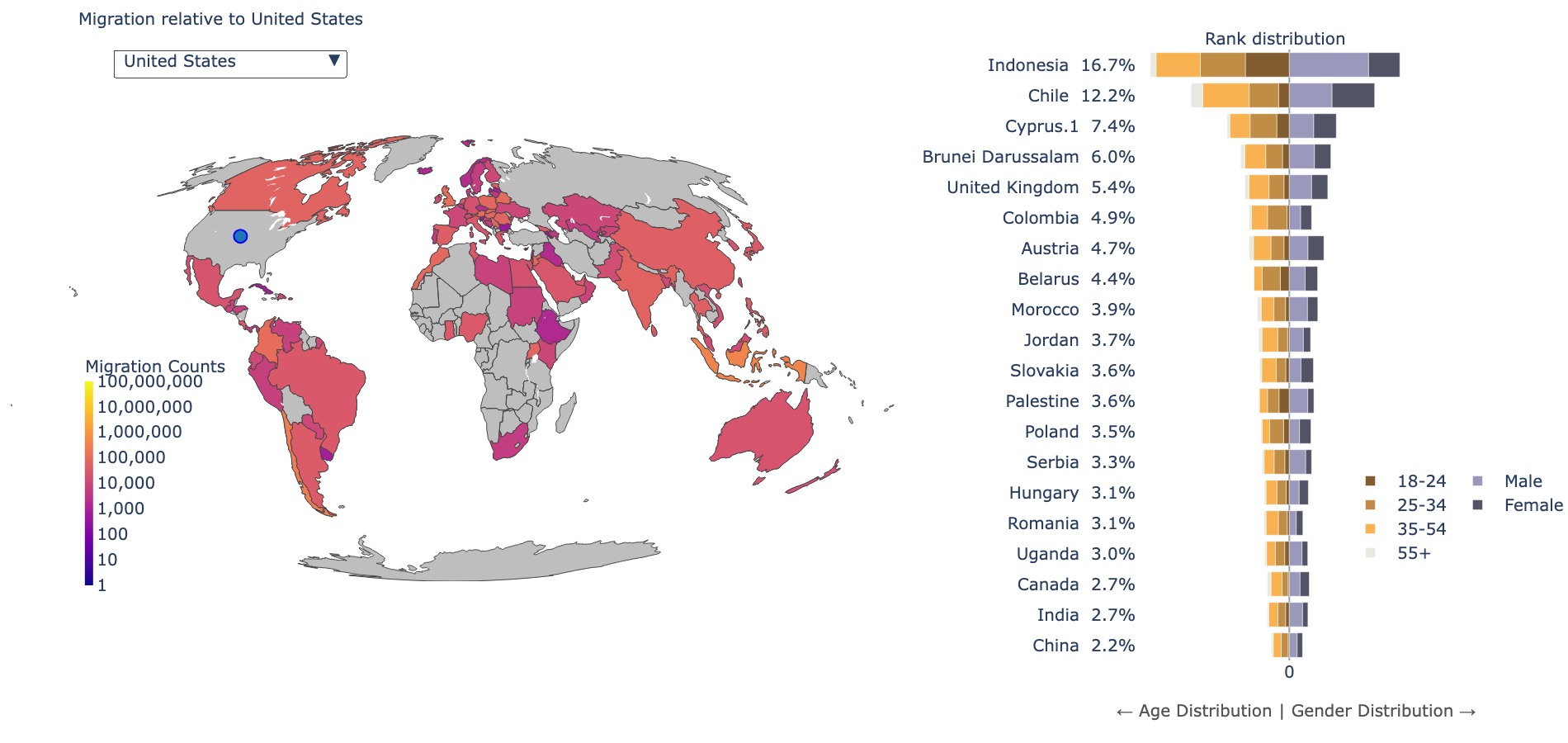}
    \caption{United States-Origin Topology and Demographic Composition of High-Potential Mobility. The map illustrates the global distribution of internationally mobile graduates from QS-ranked universities originating from the United States, with destination countries colored by total student mobility counts. The accompanying Rank Distribution (right) highlights the top 20 destination countries receiving U.S.-origin talent by relative share. Each bar in the rank distribution is disaggregated by Age Distribution (left-facing: 18--24, 25--34, 35--54, and 55+ cohorts) and Gender Distribution (right-facing: platform-inferred male and female shares), revealing outward mobility patterns that vary by destination but remain largely concentrated in prime working ages (25–54), with differing levels of gender balance across regional and global hubs.}
    \label{fig:USA}
\end{figure*}

\begin{figure*}
    \centering
    \includegraphics[width=1\linewidth]{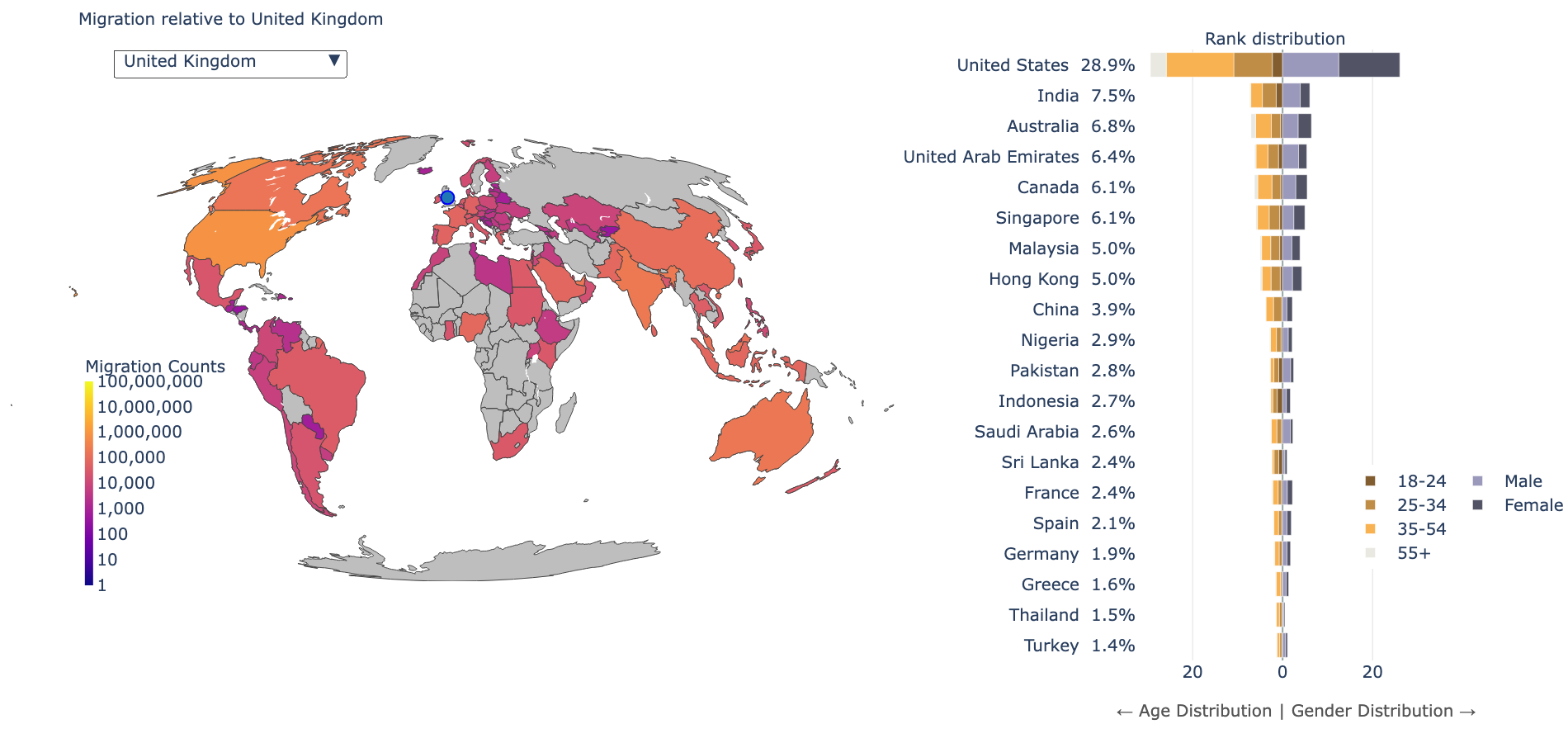}
    \caption{United Kingdom-Origin Topology and Demographic Composition of High-Potential Mobility. The map illustrates the global distribution of internationally mobile graduates from QS-ranked universities originating from the United Kingdom, with destination countries colored by total student mobility counts. The accompanying Rank Distribution (right) highlights the top 20 destination countries receiving UK-origin talent by relative share. Each bar in the rank distribution is disaggregated by Age Distribution (left-facing: 18–24, 25–34, 35–54, and 55+ cohorts) and Gender Distribution (right-facing: platform-inferred male and female shares), revealing outward mobility patterns that vary across destinations but remain predominantly concentrated in prime working ages (25–54), with observable variation in gender balance across transatlantic, Commonwealth, and regional hubs.
}
    \label{fig:uk}
\end{figure*}

\section{Composition of High-Potential Mobility}

\begin{figure*}
  \centering
  \begin{subfigure}[t]{0.32\textwidth}
    \centering
    \includegraphics[width=\linewidth]{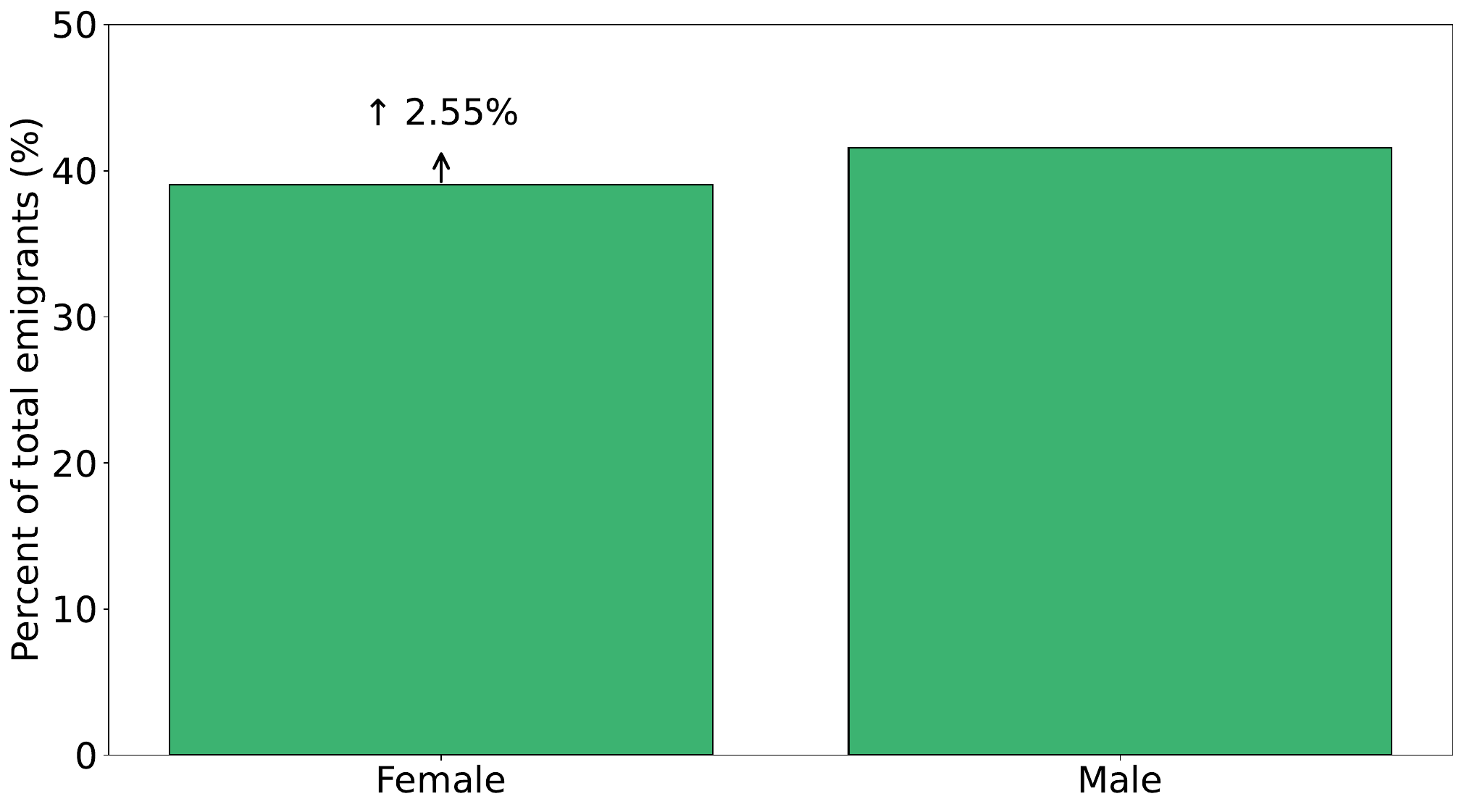}
    \caption{Gender composition of total emigrants (platform-inferred aggregates).}
    \label{fig:gender_emigration}
  \end{subfigure}\hfill
  \begin{subfigure}[t]{0.32\textwidth}
    \centering
    \includegraphics[width=\linewidth]{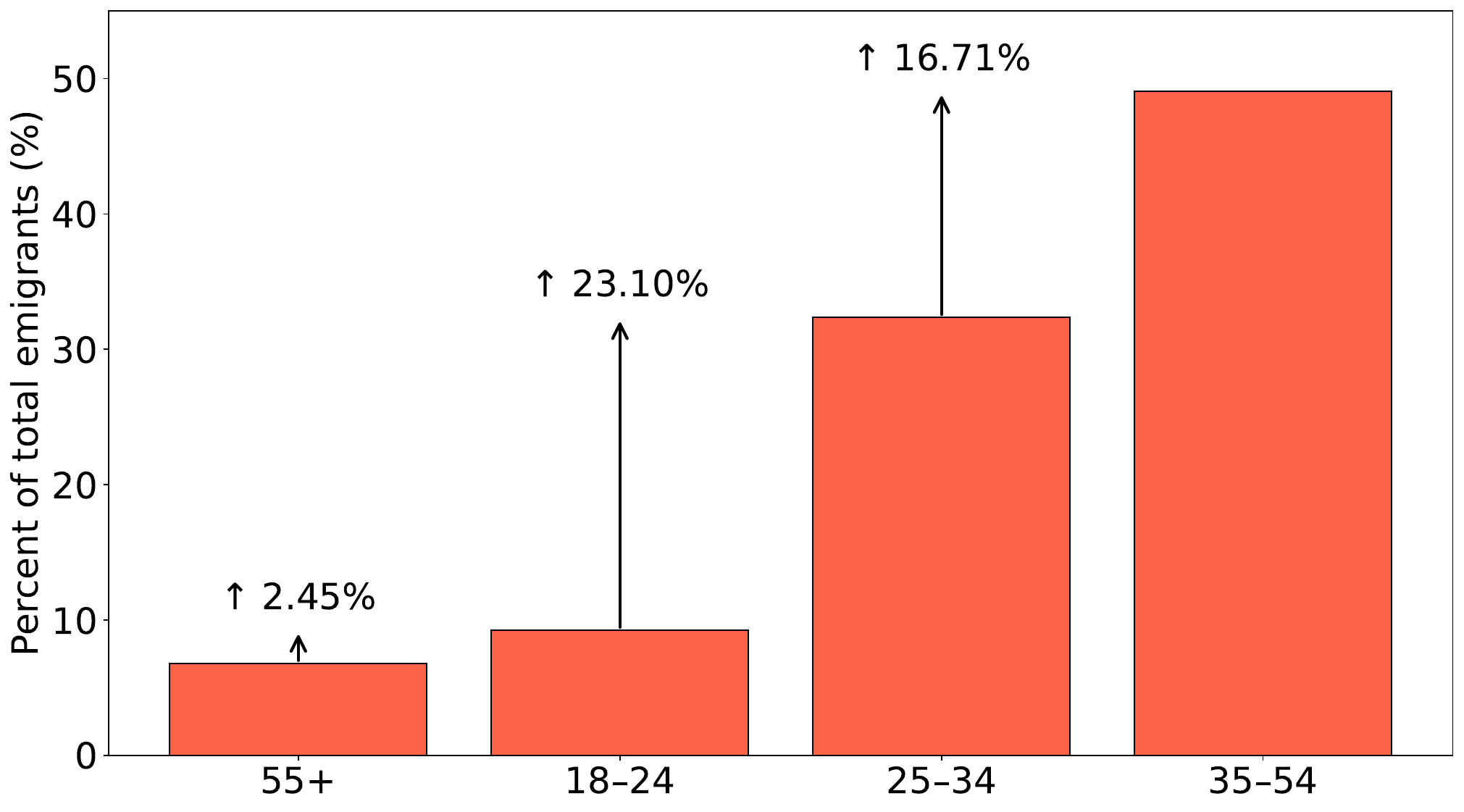}
    \caption{Age composition of total emigrants. Mobility is concentrated in prime working ages, with comparatively low representation in the 55+ group.}
    \label{fig:age_emigration}
  \end{subfigure}\hfill
  \begin{subfigure}[t]{0.32\textwidth}
    \centering
    \includegraphics[width=\linewidth]{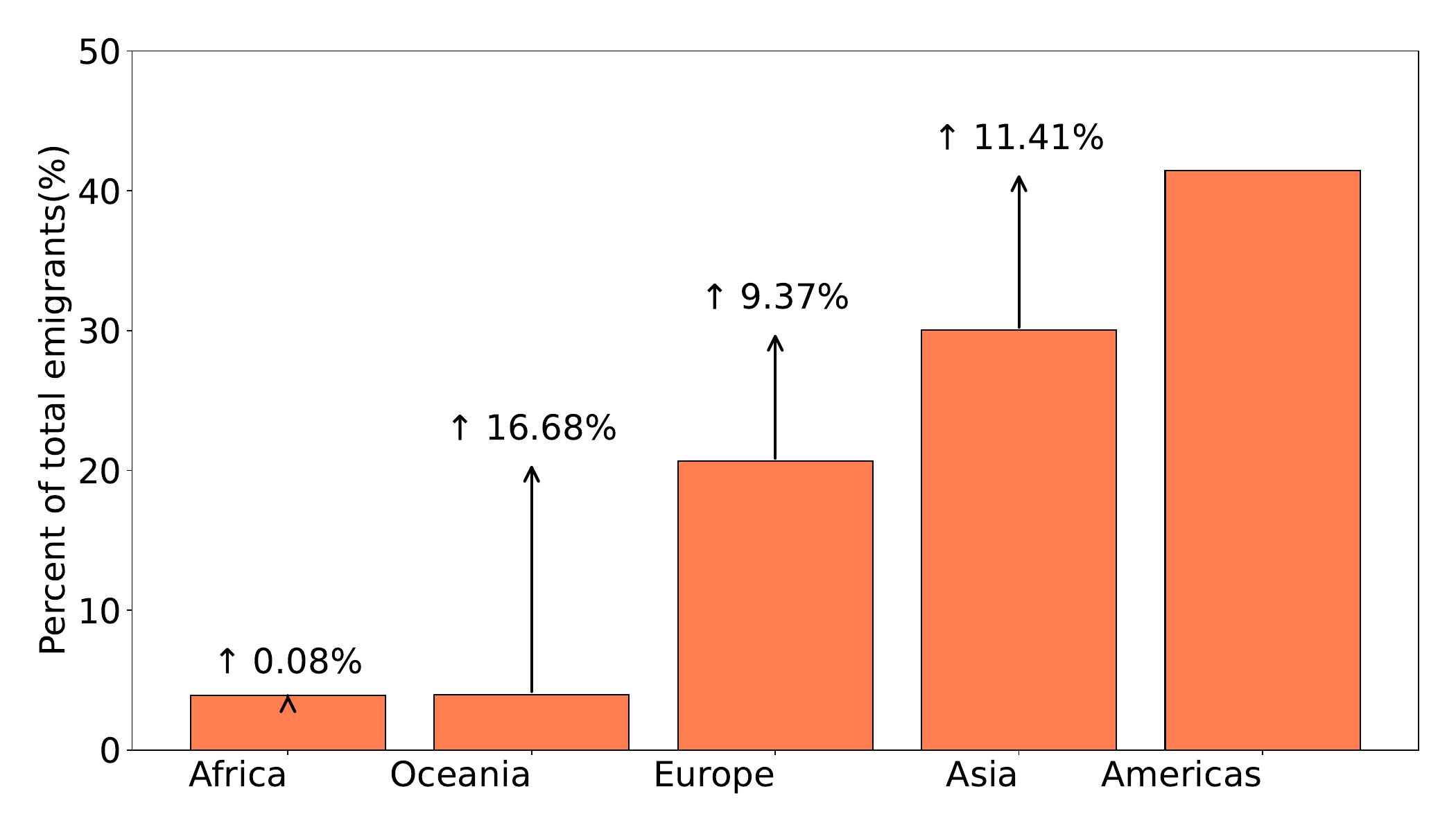}
    \caption{Regional distribution of total emigrants (percent of global total), ordered by share; arrows indicate percentage-point gaps to the next region.}
    \label{fig:region_emigration}
  \end{subfigure}
  \caption{\textbf{Figure S1. Composition of high-potential mobility.} Panels report gender, age, and regional distributions of emigrants in the LinkedIn aggregate dataset used in this study.}
  \label{fig:supp_composition}
\end{figure*}

Figure \ref{fig:supp_composition} summarizes the demographic and regional composition of high-potential mobility (QS-ranked alumni who recently relocated internationally, measured as anonymous aggregate audiences from the LinkedIn Advertising platform). Figure \ref{fig:gender_emigration} reports the gender composition of the observed emigrant population. Figure \ref{fig:age_emigration} reports the age composition, showing that mobility is concentrated in prime working ages, with the 35--54 group representing the largest share and the 55+ group the smallest. Figure \ref{fig:region_emigration} reports the regional distribution of total emigrants and visualizes stepwise gaps between ordered regional shares, highlighting pronounced differences across regions.

\section{List of Countries}
\label{sec:country}

The analysis covers 102 destination countries for which LinkedIn Campaign Manager locations were available. To avoid layout overflow in the two-column PNAS template, we report the full country list as a wrapped paragraph inside a full-width table (Table~\ref{tab:country_list}).

\begin{table}[t]
\centering
\caption{List of 102 destination countries included in the analysis.}
\label{tab:country_list}
\scriptsize
\setlength{\tabcolsep}{3pt}
\renewcommand{\arraystretch}{1.1}
\begin{tabular}{@{}p{0.33\columnwidth}p{0.33\columnwidth}p{0.33\columnwidth}@{}}
\hline
\multicolumn{3}{@{}l}{\textbf{Country list.}} \\
\hline
Argentina & Guatemala & Panama \\
Armenia & Honduras & Paraguay \\
Australia & Hong Kong SAR, China & Peru \\
Austria & Hungary & Philippines \\
Azerbaijan & Iceland & Poland \\
Bahrain & India & Portugal \\
Bangladesh & Indonesia & Puerto Rico \\
Belarus & Iraq & Qatar \\
Belgium & Ireland & Romania \\
Bosnia \& Herzegovina & Israel & Saudi Arabia \\
Brazil & Italy & Serbia \\
Brunei Darussalam & Japan & Singapore \\
Bulgaria & Jordan & Slovakia \\
Canada & Kazakhstan & Slovenia \\
Chile & Kenya & South Africa \\
China (Mainland) & Kuwait & South Korea \\
Colombia & Kyrgyzstan & Spain \\
Costa Rica & Latvia & Sri Lanka \\
Croatia & Lebanon & Sudan \\
Cuba & Libya & Sweden \\
Cyprus & Lithuania & Switzerland \\
Czechia & Luxembourg & Taiwan \\
Denmark & Macao SAR, China & Thailand \\
Dominican Republic & Malaysia & Tunisia \\
Ecuador & Malta & Türkiye \\
Egypt & Mexico & Uganda \\
Estonia & Morocco & Ukraine \\
Ethiopia & Netherlands & United Arab Emirates \\
Finland & New Zealand & United Kingdom \\
France & Nigeria & United States of America \\
Georgia & Norway & Uruguay \\
Germany & Oman & Uzbekistan \\
Ghana & Pakistan & Venezuela \\
Greece & Palestine & Viet Nam \\
\hline
\end{tabular}
\end{table}

\section{Influencing Factors Regressor}
\label{sec: Regressor}

\paragraph*{Outcome (dependent variable).}
We model student mobility volume as a log-transformed count to reduce skewness while retaining zero flows:
\[
\texttt{log\_total\_mobility}=\log\!\left(1+\texttt{total\_mobility}\right),
\]
where \texttt{total\_student\_mobility} is the observed origin-to-destination mobility count. Predictors were scaled to a common range prior to estimation (see Methods) so coefficients reflect associations per unit change on the standardized scale.

\paragraph*{Predictors used in the OLS specification.}
Table~\ref{tab:var_defs_full} defines each destination-country covariate, explains what it measures, and clarifies whether higher values are generally interpreted as better or worse and Table \ref{tab:regression_results} shows the results of regressor model for student mobility.



\begin{table*}[ht]
\centering
\caption{Definitions of destination-country predictors used in the regression.}
\label{tab:var_defs_full}
\renewcommand{\arraystretch}{1.15}
\setlength{\tabcolsep}{6pt}
\begin{tabular}{p{4.4cm} p{12.0cm}}
\toprule
\textbf{Variable} & \textbf{Definition and interpretation (destination-country)} \\
\midrule

\texttt{area} &
Total geographic area of the destination country; a proxy for physical scale and spatial capacity. Higher values indicate larger geographic scale. \\

\texttt{density} &
Population density (population per unit area), capturing crowding and pressure on infrastructure and services. Higher values typically indicate more congestion and strain. \\

\texttt{GDP\_per\_Capita} &
Gross domestic product per person, reflecting average economic resources and national capacity. Higher values indicate stronger economic capacity. \\

\texttt{Democracy} &
Democracy Index (EIU), an aggregate score across electoral process and pluralism, functioning of government, political participation, political culture, and civil liberties. Higher values indicate more democratic institutions. \\

\texttt{Global\_Gender\_Gap} &
Global Gender Gap Index (WEF), summarizing gender equality across economic participation, educational attainment, health and survival, and political empowerment; ranges from 0 (least equal) to 1 (most equal). Higher values indicate greater gender equality. \\

\texttt{Social\_Welfare\_Spending} &
Social assistance expenditure (benefits plus administrative costs) aggregated across programs and expressed as a share of GDP (most recent available year within a multi-year window). Higher values indicate greater social support effort. \\

\texttt{Poverty} &
Poverty prevalence measure (share of population below a poverty threshold, or a multidimensional poverty proxy where applicable). Higher values indicate greater poverty prevalence. \\

\texttt{Cost\_of\_Living} &
Cost of living index reflecting relative prices of goods, services, and housing. Higher values indicate higher living costs and lower affordability. \\

\texttt{Economic\_Freedom} &
Economic freedom index capturing market openness and regulatory constraints, typically incorporating elements of rule of law, government size, regulatory efficiency, and open markets. Higher values indicate greater economic freedom. \\

\texttt{Remittances} &
Remittances received, defined as non-commercial transfers from workers abroad to households with family ties, representing a key channel of international capital flows and diaspora ties. Higher values indicate stronger remittance inflows. \\

\texttt{Unemployment\_Rate} &
Unemployment rate, capturing the share of the labor force without work but actively seeking employment (estimates may vary by source and methodology). Higher values indicate weaker labor-market conditions. \\

\texttt{Education} &
Education environment proxy capturing educational attainment and access (index-style measure drawn from global development reporting), reflecting the strength of the local education ecosystem. Higher values indicate stronger education conditions. \\

\texttt{Life\_Expectancy} &
Life expectancy at birth (years), a summary indicator of population health and overall living conditions. Higher values indicate better health outcomes. \\

\texttt{Healthcare\_Expenditure} &
Healthcare spending indicator reflecting national investment in health services; interpretation may vary because higher spending can reflect greater investment or higher system costs. Higher values indicate more healthcare spending. \\

\texttt{Climate\_Change\_Performance} &
Climate Change Performance Index style composite summarizing greenhouse gas emissions, energy use, renewable energy transition, and climate policy strength. Higher values indicate stronger climate performance. \\

\texttt{Women\_Peace\_and\_Security} &
Women, peace, and security index capturing women’s inclusion, justice, and security. Higher values indicate better women’s security and inclusion. \\

\texttt{Global\_Peace} &
Peace and safety composite capturing societal peacefulness and stability. Higher values indicate more peace and stability. \\

\bottomrule
\end{tabular}
\end{table*}

\begin{table*}
\centering
\caption{OLS Regression Results for Log Total Student Mobility}
\label{tab:regression_results}
\begin{tabular}{l c c c c}
\toprule
\textbf{Variable} & \textbf{Coefficient} & \textbf{Std. Error} & \textbf{\textit{t}-statistic} & \textbf{\textit{P} $> |t|$} \\
\midrule
Density & -1.6285 & 0.330 & -4.934 & 0.000 \\
Area & 4.1479 & 0.179 & 23.153 & 0.000 \\
GDP per Capita & 1.8706 & 0.348 & 5.373 & 0.000 \\
Democracy & 1.1055 & 0.210 & 5.261 & 0.000 \\
Global Gender Gap & -0.6090 & 0.303 & -2.008 & 0.045 \\
Social Welfare Spending & -0.2520 & 0.303 & -0.832 & 0.406 \\
Poverty & -0.2979 & 0.225 & -1.324 & 0.185 \\
Cost of Living & 0.6038 & 0.392 & 1.539 & 0.124 \\
Economic Freedom & 0.6388 & 0.248 & 2.577 & 0.010 \\
Remittances & 1.0732 & 0.206 & 5.198 & 0.000 \\
Unemployment Rate & -0.6846 & 0.276 & -2.477 & 0.013 \\
Education & 0.9474 & 0.375 & 2.523 & 0.012 \\
Life Expectancy & -0.4806 & 0.307 & -1.566 & 0.117 \\
Healthcare Expenditure & -0.5430 & 0.227 & -2.390 & 0.017 \\
Climate Change Performance & 0.0674 & 0.203 & 0.332 & 0.740 \\
Women, Peace, and Security & -0.1046 & 0.384 & -0.272 & 0.785 \\
Global Peace & 1.1536 & 0.246 & 4.685 & 0.000 \\
\midrule
\textbf{Model Diagnostics} & & & & \\
Observations & 10,403 & & $R^{2}$ (uncentered) & 0.336 \\
$F$-statistic (17, 10386) & 308.9 & & Adj. $R^{2}$ (uncentered) & 0.335 \\
Prob ($F$-statistic) & 0.000 & & Log-Likelihood & -26,786 \\
AIC & 5.361e+04 & & BIC & 5.373e+04 \\
\bottomrule
\end{tabular}

\end{table*}

\section{Multivariate Regression Analysis of Gender-Disaggregated Mobility}
\label{sec:gender_regression}

To further elucidate the drivers of high-skilled mobility, we present the detailed results of our multivariate Ordinary Least Squares (OLS) regression models for male and female migrant flows. To stabilize variance while retaining zero-flow observations, the dependent variables were transformed using the natural logarithm as follows:$$\texttt{log\_male\_mobility} = \log\!\left(1+\texttt{male\_mobility}\right),$$$$\texttt{log\_female\_mobility} = \log\!\left(1+\texttt{female\_mobility}\right).$$These models assess the impact of 17 socioeconomic and demographic predictors across 10,403 observations. The results, summarized in Table \ref{tab:gender_regression}, demonstrate that while structural drivers such as destination Area, GDP per Capita, Democracy, and Remittances remain significant positive "pull'' factors for both cohorts, their relative influence varies. Notably, female mobility exhibits a statistically significant and stronger sensitivity to the Cost of Living ($\beta = 1.05$) compared to male mobility ($\beta = 0.62$, $p = 0.06$), while Economic Freedom serves as a marginally significant driver for males only.

\begin{table*}
\centering
\caption{Comparative Regression Results for Male and Female Migrants}
\label{tab:gender_regression}
\begin{tabular}{l c c c c}
\toprule
& \multicolumn{2}{c}{\textbf{Male Migrants}} & \multicolumn{2}{c}{\textbf{Female Migrants}} \\
\cmidrule(r){2-3} \cmidrule(l){4-5}
\textbf{Variable} & \textbf{Coefficient} & \textbf{P $> |t|$} & \textbf{Coefficient} & \textbf{P $> |t|$} \\
\midrule
Density & -0.9720 & 0.000 & -0.6586 & 0.013 \\
Area & 3.4478 & 0.000 & 3.2007 & 0.000 \\
GDP per Capita & 1.4233 & 0.000 & 0.8960 & 0.001 \\
Democracy & 0.8340 & 0.000 & 0.7983 & 0.000 \\
Global Gender Gap & -0.7275 & 0.004 & -0.6439 & 0.008 \\
Social Welfare Spending & -0.1658 & 0.516 & -0.0262 & 0.914 \\
Poverty & -0.1474 & 0.437 & -0.4357 & 0.016 \\
Cost of Living & 0.6211 & 0.060 & 1.0491 & 0.001 \\
Economic Freedom & 0.3974 & 0.057 & 0.1922 & 0.334 \\
Remittances & 0.9323 & 0.000 & 0.8132 & 0.000 \\
Unemployment Rate & -0.6749 & 0.004 & -0.6777 & 0.002 \\
Education & 0.7847 & 0.013 & 0.8954 & 0.003 \\
Life Expectancy & -0.0761 & 0.769 & -0.2246 & 0.361 \\
Healthcare Expenditure & -0.1834 & 0.338 & 0.3216 & 0.078 \\
Climate Change Performance & -0.0102 & 0.952 & -0.0713 & 0.661 \\
Women, Peace, and Security & -0.6315 & 0.051 & -0.4809 & 0.118 \\
Global Peace & 0.6911 & 0.001 & 0.3579 & 0.070 \\
\midrule
\textbf{Diagnostics} & & & & \\
Observations & 10,403 & & 10,403 & \\
$R^2$ (uncentered) & 0.252 & & 0.240 & \\
$F$-statistic & 205.9 & & 193.2 & \\
Log-Likelihood & -25,011 & & -24,487 & \\
\bottomrule
\end{tabular}
\end{table*}

\begin{figure*}
    \centering
    \includegraphics[width=0.75\linewidth]{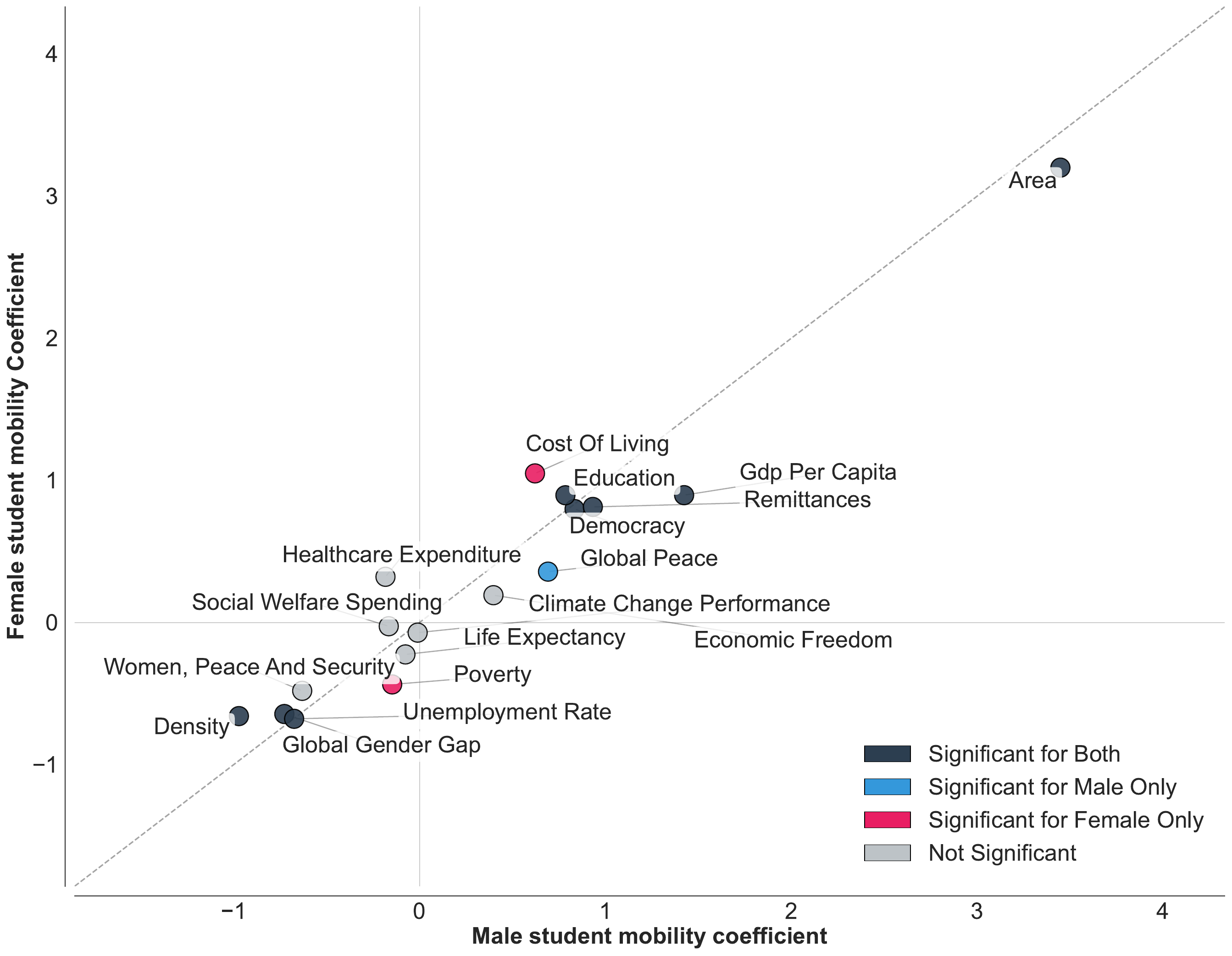}
    \caption{\textbf{Comparative drivers of male versus female high-potential mobility.}
Scatterplot of OLS coefficient estimates from gender-disaggregated models across 102 destination countries (male: $R^2=0.252$; female: $R^2=0.240$), with male coefficients on the horizontal axis and female coefficients on the vertical axis. Points are colored by significance at $p<0.05$ (significant for both models; male only; female only; not significant). The dashed line ($y=x$) denotes equal associations for men and women, and the vertical/horizontal reference lines mark zero effects.}

    \label{fig:gender_cofficient}
\end{figure*}

\end{document}